\documentclass[11pt]{article}

\usepackage[a4paper,margin=1in]{geometry}
\usepackage[T1]{fontenc}
\usepackage{lmodern}
\usepackage{microtype}
\usepackage{graphicx}
\graphicspath{{figs/}}
\usepackage{booktabs}
\usepackage{amsmath,amssymb}
\usepackage{xcolor}
\usepackage{siunitx}
\usepackage{tikz}                 %
\usetikzlibrary{arrows.meta,calc,fit}
\usepackage{algorithm}
\usepackage{algpseudocode}
\usepackage{float}                %
\usepackage[numbers,sort&compress]{natbib}
\usepackage[hidelinks]{hyperref}

\usepackage{caption}
\usepackage[capitalise,noabbrev]{cleveref}
\crefformat{appendix}{#2#1#3}
\Crefformat{appendix}{#2#1#3}
\crefrangeformat{appendix}{#3#1#4--#5#2#6}
\crefmultiformat{appendix}{#2#1#3}{ and~#2#1#3}{, #2#1#3}{ and~#2#1#3}

\newfloat{listing}{tbp}{lol}
\floatname{listing}{Listing}
\crefname{listing}{Listing}{Listings}
\Crefname{listing}{Listing}{Listings}

\newcommand{\gradsolve}{{\normalfont\scshape gradsolve}}
\newcommand{\diffrax}{Diffrax}
\newcommand{\code}[1]{\texttt{#1}}
\newcommand{\recordreplay}{record-and-replay}
\newcommand{\x}{\ensuremath{\times}}

\newcommand{\revwprange}{7.8--9.3\x}

\newcommand{\fourSolverDiffrax}{8.8--9.8\x}

\newcommand{\revDenseRange}{9.0--10.7\x}

\newcommand{\fdNoiseFloor}{\num{7e-9}}

\newcommand{\torchodeRange}{58--63\x}

\newcommand{\torchdiffeqRange}{83--99\x}
\newcommand{\torchodeCompiledRange}{26--29\x}

\newcommand{\stiffRange}{0.8--5.6\x}

\newcommand{\stiffRangeMax}{5.6\x}

\newcommand{\stiffOverlap}{2.2}

\newcommand{\mevVernLorenz}{9.5\x}
\newcommand{\mevRodasRobertson}{7.2\x}
\newcommand{\mevRodasHires}{12.9\x}
\newcommand{\mevVernLinear}{43.3\x}

\newcommand{\gradAccLorenz}{\num{4.6e-06}}
\newcommand{\gradAccRobertson}{\num{6.8e-06}}
\newcommand{\sharedMeshLorenz}{\num{2.9e-15}}
\newcommand{\sharedMeshVdp}{\num{1.8e-13}}
\newcommand{\errMedLorenz}{\num{4.8e-07}}
\newcommand{\errMaxLorenz}{\num{7.6e-06}}
\newcommand{\errMedRobertson}{\num{3.2e-05}}
\newcommand{\errMaxRobertson}{\num{4.4e-05}}

\newcommand{\frozenMeshRatios}{10.7\x, 7.9\x, and 8.3\x}

\newcommand{\crossarchRange}{5.6--14.1\x}

\newcommand{\crossarchFourNinety}{5.6--12.5\x}

\newcommand{\fitLorenz}{5.39--6.32\x}

\newcommand{\fitGalactic}{3.97--4.50\x}

\newcommand{\fitRobertson}{1.14--1.35\x}

\newcommand{\fwdLaneThroughput}{\SI{0.0038}{\micro\second}/trajectory}

\newcommand{\dispersionRange}{1.33--1.34}

\newcommand{\revVdPRange}{9.7--10.5\x}

\newcommand{\hiresLowOrder}{4.1--4.3\x}

\newcommand{\memRecordVsTape}{about \SI{10}{\mega\byte}; the JAX replay of the same
mesh, storing every intermediate value for its backward sweep, requested
\SI{24.3}{\giga\byte} and failed to obtain it on a \SI{24}{\giga\byte} card}

\newcommand{\stiffSecondArch}{2.5--2.8\x}

\newcommand{\stiffSweepRange}{2.4--3.4\x}

\newcommand{\cpuGpuCrossoverTop}{43.8\x}

\newcommand{\udeFdFused}{\num{1.4e-7}}

\newcommand{\udeFdReplay}{\num{1.55e-6}}

\newcommand{\fusedWpRange}{2.2--2.4\x}

\newcommand{\steptoLoopFormGpu}{7.9--11.1\x}
\newcommand{\scanWpRange}{1.9--2.6\x}

\newcommand{\fwdSamePodDouble}{2.8\x}

\newcommand{\fwdSamePodSingle}{1.95\x}

\newcommand{\fwdDiffraxPoint}{54\x}

\newcommand{\scaleAxisTop}{\SI{0.00248}{\micro\second}/trajectory at $n = 2^{24}$}

\newcommand{\capacityDemo}{$10^{8}$ trajectories in \SI{0.25}{\second}}

\newcommand{\reverseSweepTop}{\num{1048576}}

\newcommand{\reverseLargeNRange}{6.2--8.5\x}

\newcommand{\lorenzChaoticPoint}{6.5\x}

\newcommand{\fwdModeLorenzRange}{3.2--3.5\x}      %
\newcommand{\fwdReplayLorenzRange}{7.3--7.9\x}    %
\newcommand{\dfxFwdOverRevLorenz}{2.6--2.7\x}     %
\newcommand{\fwdModeRobertsonRange}{0.3--2.2\x}   %
\newcommand{\fwdReplayRobertsonRange}{0.4--3.0\x} %
\newcommand{\dfxFwdOverRevRobertson}{2.5\x}       %
\newcommand{\fwdModeLorenzSweep}{2.4--4.0\x}      %
\newcommand{\fwdReplayLorenzSweep}{3.7--10.1\x}   %
\newcommand{\fwdModeVdPRange}{3.3--3.7\x}         %
\newcommand{\fwdReplayVdPRange}{10.7--13.2\x}     %
\newcommand{\fwdModeRobertsonSweep}{1.2\x}        %
\newcommand{\fwdReplayRobertsonSweep}{1.5--1.6\x} %
\newcommand{\fwdModeVernLorenz}{3.0\x}            %
\newcommand{\fwdReplayVernLorenz}{6.6\x}          %
\newcommand{\fwdModeRodasRobertson}{2.8\x}        %
\newcommand{\fwdReplayRodasRobertson}{3.3\x}      %

\newcommand{\recordDeviceLorenz}{\SI{0.0076}{\second}}     %
\newcommand{\recordDeviceRobertson}{\SI{0.36}{\second}}    %
\newcommand{\recordDeviceHires}{\SI{0.7}{\second}}         %
\newcommand{\recordDeviceUserLorenz}{\SI{0.043}{\second}}  %
\newcommand{\recordPaybackRange}{0.23--0.97}               %

\newcommand{\timingSpread}{12.6\%}

\newcommand{\cpuLaptop}{Apple M3 Max}
\newcommand{\cpuPodHost}{AMD EPYC 7742 64-Core Processor}
\newcommand{\cpuPodHostB}{Intel Xeon Platinum 8470}

\newcommand{\fitGalacticSeconds}{21\,s with \gradsolve{} against 94\,s with \diffrax{}}

\newcommand{\fitPinnedMargin}{6.6\%}

\newcommand{\udeSmallBatch}{2.0\x}
\newcommand{\udeLargeBatch}{1.3\x}

\newcommand{\singleLorenzGradRatio}{0.52}
\newcommand{\singleRobertsonGradRange}{1.00--1.04}
\newcommand{\singleGradAgreement}{\num{1.8e-6}}

\newcommand{\stiffFwdWarpRange}{1.5--4.5\x}
\newcommand{\cudaRosenHiresVsDiffeqgpu}{1.5--3.1\x}
\newcommand{\cudaRosenHiresVsWarp}{1.8--2.3\x}
\newcommand{\cudaRosenRobertsonSlow}{1.1--5.3\x}
\newcommand{\divFwdRange}{1.4--2.0\x}

\title{\gradsolve{}: fast exact gradients for ODE ensembles on GPUs}
\author{%
  Alessio Spurio Mancini\\[2pt]
  \small Department of Physical Sciences and Engineering,\\
  \small Royal Holloway, University of London, Egham Hill, Egham TW20 0EX, United Kingdom\\
  \small \texttt{alessio.spuriomancini@rhul.ac.uk}%
}
\date{}

\begin{document}

\maketitle

\begin{abstract}
Ordinary differential equations (ODEs) underlie models in science and
engineering, and many applications need derivatives of their solutions with
respect to parameters. Ensembles of independent trajectories suit graphics
processing units (GPUs), but current GPU software forces a trade-off: the fastest
ensemble solvers cannot be differentiated in reverse mode at the speed they solve,
and the solvers built for differentiation solve more slowly. No single tool has
yet offered a reverse-mode gradient at the speed of a fused-kernel solve.

We present \gradsolve{}, an open-source JAX library for solving and reverse-mode
differentiating low-dimensional ODE ensembles on NVIDIA GPUs. It records the steps
an adaptive solver accepts and differentiates a fixed-step replay of them; the
returned gradient is the exact discrete adjoint of those steps, the same
derivative \diffrax{} returns by default, obtained more cheaply from a
fixed-length chain than from an adaptive loop. It targets ensembles differentiated
many times against one recorded mesh, keeps \diffrax{} as a fallback, and supports
explicit and Rosenbrock integrators.

Used as a solver, \gradsolve{}'s forward-only kernel ran \fwdSamePodDouble{}
faster than DiffEqGPU.jl; used for gradients, once a record exists, it computed
them \crossarchRange{} faster than \diffrax{}'s checkpointed adjoint at matched
forward-state accuracy across three GPU generations, the advantage narrowing on
large ensembles and, on stiff systems, down to parity at tight accuracy.
\gradsolve{} is released at \url{https://github.com/ECLIPSE-AI4Science/gradsolve}.


\end{abstract}

\section{Introduction}\label{sec:intro}

Ordinary differential equations (ODEs) underpin quantitative modeling
throughout science and engineering, and the equations of practical
interest only rarely admit closed-form solutions
\citep{hairer1993solving, hairer1996solving}. Their numerical integration
is therefore a routine component of scientific computing, supported by mature
solver suites in every major ecosystem
\citep{hindmarsh2005sundials, shampine1997matlab, virtanen2020scipy,
rackauckas2017differentialequations}, and the solver's
efficiency frequently governs the overall cost of the computation it is
embedded in, particularly when the same equation must be solved many times. Alongside raw solver
efficiency, a second requirement has become equally central: the ability
to differentiate the solution of an ODE with respect to its parameters.
Such gradients are required in several settings. In the training of neural
differential equations, models in which a neural network forms part of the
right-hand side of the equation, a loss must be differentiated through the
numerical integrator at each optimization step
\citep{chen2018neuralode, rackauckas2020universal, kidger2022neural}. In the
calibration of mechanistic models, rate constants, initial conditions, or
forcing terms are adjusted iteratively until the simulated trajectory
reproduces observed data \citep{raue2009identifiability, tarantola2005inverse}.
Sensitivity analysis quantifies the response of the solution to perturbations
of the inputs \citep{cao2003adjoint, serban2005cvodes, alexe2009forward}, and
optimal control seeks the control trajectory minimizing a prescribed cost
functional \citep{giles2000adjoint, betts2010practical}.
Each of these tasks requires the same quantity, the gradient of an ODE
solution with respect to its parameters, and whenever the parameters
outnumber the outputs, as is typical, reverse-mode differentiation, which
obtains the whole gradient from one backward sweep through the computation, needs the fewest passes.

In practice, such gradients are seldom required for a single trajectory
in isolation. Calibration procedures, for example, evaluate many
parameter settings simultaneously; each training step differentiates a loss through a batch of trajectories; and uncertainty quantification
propagates thousands of parameter draws through the same equation
\citep{smith2013uncertainty}.
The resulting workload, an ensemble of independent ODE solves, is
naturally parallel and therefore well suited to graphics processing
units (GPUs). On the GPU,
however, computational speed and differentiability have so far been
provided by disjoint families of tools. The fastest ensemble solvers
compile the entire adaptive integration procedure, including step-size
control and error estimation, into a single GPU kernel, the program the GPU runs.
Each of the GPU's many parallel threads, its smallest units of execution,
integrates one trajectory; such a kernel is called \emph{fused} below. DiffEqGPU.jl
\citep{utkarsh2024diffeqgpu, rackauckas2017differentialequations}
represents the published state of the art, alongside the CUDA solver suite
MPGOS \citep{nagy2022mpgos}. The compiled control flow that makes these kernels fast, however, cannot be
traced by the reverse-mode automatic differentiation of the language they are
written in. Reverse-mode gradients are offered only through a slower path that
advances the whole ensemble together as one array, and, to our knowledge, no
reverse-mode ensemble gradient has been reported that keeps the cost advantage
of these kernels over the array-based solvers.
Conversely, the solvers designed for differentiation, \diffrax{}
\citep{kidger2021diffrax, kidger2022neural} in JAX \citep{bradbury2018jax} and
torchode \citep{lienen2022torchode} in PyTorch \citep{paszke2019pytorch},
differentiate the adaptive loop itself. Even where the accepted step sizes are held fixed, the program being differentiated is still the loop that chose them, with its error tests and branches and, in a compiled framework such as JAX, a fixed upper bound on its number of iterations. Its backward sweep is correspondingly more expensive than one over a fixed-length chain. torchdiffeq \citep{chen2018torchdiffeq}, also in PyTorch, offers both routes;
the interface we benchmark, \code{odeint\_adjoint}, avoids differentiating the loop by solving a second equation backward in time for the sensitivities (a continuous adjoint), at the price of a gradient that is not the exact derivative of the solution the forward solver returned. Two further tools are not timed here. The solver in JAX's own experimental module \citep{bradbury2018jax} returns a continuous adjoint; the Julia sensitivity suite \citep{ma2021comparison} returns a discrete adjoint, the derivative of the steps actually taken, but only on its array-based paths. Neither provides an adaptive per-trajectory GPU ensemble solve with a discrete adjoint. Fast solving
and differentiability have moreover sat in different tools and largely in different
language ecosystems: the fused kernels are written in Julia and CUDA C++,
whereas Python, where most differentiable modeling is done, has had only the
differentiable solvers. A practitioner working in Python has therefore had
differentiable solvers but no fused ensemble solver, and in no ecosystem has
the fastest ensemble solver also been the one that returns a reverse-mode
gradient.

\gradsolve{} is an open-source GPU library, written in Python and built on JAX,
designed to eliminate this trade-off and released at
\url{https://github.com/ECLIPSE-AI4Science/gradsolve}. To differentiate an ensemble, it first solves each trajectory once with a
standard adaptive integrator and records the sequence of steps the integrator
accepted. It then computes the gradient by differentiating a fixed-step
\emph{replay} of the recorded steps, compiled as one regular program over the
whole ensemble. The right-hand side is an ordinary JAX function written by the user. Its
recording runs as a batched JAX loop on the GPU, or, on request, the library translates the function into a single fused GPU kernel, the form in which the built-in systems run (\cref{sec:method-interface}). Because the steps of the replay are
predetermined, the differentiated computation is a fixed-length chain of
identical steps with no branch in it. Its backward sweep is one linear pass
back through the chain, whereas an adaptive loop differentiated in place must be given a fixed upper
bound on its length, must decide which intermediate states to store and which
to recompute (checkpointing), and is carried through every trial of every
trajectory. This difference in the form of the backward sweep, not in the kind of
derivative computed, is where \cref{sec:whyfast} traces the speedups reported below: holding the method and the mesh fixed, the fixed-step
scan is several times cheaper than \diffrax{}'s loop on the GPU. The
returned gradient is exact in a precisely defined sense: it is the exact derivative of the replay, with the recorded steps held fixed as
data, rather than a derivative of the adaptive solve that generated them.
When no gradient is required, the same interface sends those systems to the forward-only kernel, so that no cost is incurred for differentiability
when it is not used. The derivative \gradsolve{} returns is classical:
differentiating a previously accepted step sequence is the
discrete-adjoint construction developed for sensitivity analysis and
PDE-constrained optimization
\citep{hager2000runge, sandu2006properties, zhang2014fatode,
farrell2013automated, zhang2022tsadjoint, gholami2019anode, onken2020discretize,
kidger2022neural}; \cref{sec:method} places it among its neighbors. To our knowledge, \gradsolve{} is the first implementation of this construction
whose record is produced by a fused GPU ensemble kernel and whose backward sweep
is a single fixed-length sweep over the whole ensemble. The workloads \gradsolve{} is built for include ensemble calibrations and
sensitivity studies of low-dimensional mechanistic models and the training of
small neural right-hand sides: cases in which the same ensemble is differentiated many times against one
recorded set of steps.

Among the differentiable solvers, \diffrax{} is the fastest differentiable baseline
we measured. Its default adjoint, called the checkpointed adjoint below,
differentiates the steps the solver actually
took, exactly, and saves memory by storing only some intermediate states and
recomputing the rest; the ensemble runs as a single compiled array program. \gradsolve{} returns the same discrete adjoint as \diffrax{}, specialized to the regime of low-dimensional GPU ensembles differentiated many times against one record, and keeps \diffrax{} itself as a routing fallback. Of the PyTorch pair, torchode assigns each trajectory an independent
step-size controller, and torchdiffeq returns the continuous-adjoint gradient
described above. We benchmark against all three, and
against DiffEqGPU.jl on the forward side. At matched achieved accuracy, that is, with each baseline's tolerance adjusted
until its error equals ours, \gradsolve{} computes gradients \revwprange{}
faster than \diffrax{}'s checkpointed adjoint on a Lorenz ensemble. Across A100, H100, and RTX~4090 GPUs the margin is \crossarchRange{}, declining with ensemble size on every card. These ranges differ because the first is a tolerance sweep at one ensemble size and the
second spans ensemble size and card; the figure to carry away is that the gradient costs several
times less than \diffrax{}'s, the saving largest on small non-stiff ensembles. Against \diffrax{}'s forward mode, the cheaper setting when each trajectory carries one to three inputs, the margin is \fwdModeLorenzRange{} for the replay in reverse mode and \fwdReplayLorenzRange{} for the replay in forward mode. It is faster on every system we measured except on the stiff Robertson system at the tighter accuracies, where \diffrax{}'s higher-order stiff method reaches a given accuracy in fewer steps than \gradsolve{}'s default second-order one. The advantage is therefore largest on non-stiff ensembles and on stiff ones short of the very tightest accuracy. Against the PyTorch solvers the gap widens to between one
and two orders of magnitude, consistent with the cost, measured below in their
forward passes alone, of launching each solver operation as its own GPU kernel from the CPU that
drives the GPU. On the forward side,
\gradsolve{}'s forward-only kernel runs \fwdSamePodDouble{} faster than
DiffEqGPU.jl's fused kernel on the same GPU in double precision, and
\fwdSamePodSingle{} faster in single precision (\cref{sec:benchmarks-forward}).

\looseness=-1 The remainder of the paper is organized as follows. \Cref{sec:diff}
formulates the problem of differentiating an adaptive ensemble solve.
\Cref{sec:method} presents the record-and-replay method and the four
integrators it covers. \Cref{sec:benchmarks} reports the forward and gradient
benchmarks against DiffEqGPU.jl, \diffrax{}, torchode, and torchdiffeq,
together with three parameter-estimation examples in which the lower gradient
cost yields a faster fit at matched accuracy; \ref{app:models} defines the
benchmark problems. \Cref{sec:whyfast} identifies the origin of the measured
advantage through controlled experiments and states when to expect it. \Cref{sec:discussion} summarizes the results and their range
of validity, and concludes; \ref{app:repro} gives the benchmark protocol and the gradient checks.

\section{Differentiating an ODE solve}\label{sec:diff}

The gradient to be computed is that of a scalar loss $L$, formed from
many solutions of the same equation, with respect to the parameters
$\theta$. This section explains why that gradient is hard to compute
quickly on a GPU. Throughout, an ensemble is a collection of $n$ trajectories
of the same system $\dot y = f(t, y, \theta)$, $y \in \mathbb{R}^d$, solved at once from $n$ initial conditions, with parameters $\theta$ that may
be shared or vary from trajectory to trajectory.

\subsection{Adaptive Runge--Kutta solvers}\label{sec:diff-adaptive}

An adaptive solver takes each step as large as accuracy allows: long
steps where the solution is smooth, short ones where it changes quickly.
Because the right size is not known in advance, it proceeds by trial. From
the state $(t, y)$ it attempts a step of size $h$ with an explicit
Runge--Kutta method \citep{hairer1993solving}. Such a method evaluates the
right-hand side $s$ times per step, or one fewer per accepted step when the
last stage can be reused as the first stage of the next step. Each evaluation, a \emph{stage} $k_j$,
samples the slope at an intermediate state built from the stages before it,
\begin{equation}
  k_j \;=\; f\Bigl(t + c_j h,\;\, y + h \textstyle\sum_{l<j} a_{jl}\, k_l,\;\,
  \theta\Bigr), \qquad j = 1, \dots, s,
  \label{eq:stages}
\end{equation}
with fixed coefficients $a_{jl}$ and $c_j$. Fixed weights $b_j$ then complete
the step,
\begin{equation}
  y_{\mathrm{new}} \;=\; y + h \textstyle\sum_{j=1}^{s} b_j\, k_j .
  \label{eq:update}
\end{equation}
An \emph{embedded} pair carries a second set of
weights $\hat b_j$ combining the same $s$ evaluations into a lower-order
$\hat y_{\mathrm{new}}$, so $e = y_{\mathrm{new}} - \hat y_{\mathrm{new}}$
estimates the step's local error without further evaluations of $f$
\citep{dormand1980family}. The step is accepted when the root-mean-square of the componentwise ratio $e /
(\tau_{\mathrm{abs}} + \tau_{\mathrm{rel}} \max(\lvert y\rvert, \lvert
y_{\mathrm{new}}\rvert))$ is at most one
\citep{hairer1993solving, shampine1997matlab}: the absolute tolerance
$\tau_{\mathrm{abs}}$ keeps the test meaningful where the state is near zero,
the relative tolerance $\tau_{\mathrm{rel}}$ scales it with the state. If
the test passes, $t$ and $y$ advance and a controller
proposes the next step size from the error estimate
\citep{soderlind2002automatic}; if it fails, the step is
retried from the same $(t, y)$ with a smaller $h$.
Which step \gradsolve{} takes depends on whether the user declares the problem
\emph{stiff}, that is, whether it couples fast-decaying components to the
slow dynamics of interest. An explicit method must then take very small steps
to remain stable even where the solution is smooth, while an implicit or
linearly implicit step stays stable at step sizes set by accuracy alone.
\gradsolve{}'s default for non-stiff systems is the explicit fifth-order pair of \citet{tsitouras2011runge}, called Tsit5 below, with $s = 7$ stages, of which the last is reused as the first stage of the next accepted step. For stiff systems the default is a linearly implicit Rosenbrock--Wanner step, which solves one linear system involving the Jacobian of $f$ per stage \citep{rosenbrock1963general, shampine1997matlab}. \Cref{sec:method} describes the specific methods, including higher-order stiff and non-stiff variants.

Two properties of this loop matter in what follows. The number of
steps a trajectory needs is discovered at run time, so
two trajectories of the same ensemble, integrated to the same tolerance, can
accept step sequences of different lengths. And the accepted steps are not
all the work: reaching the end also involved rejected trials, error
estimates, and controller updates.

\subsection{Forward-mode and reverse-mode differentiation}\label{sec:diff-admodes}

The quantity we need is the gradient $\nabla_\theta L$ of a scalar loss $L$
with respect to the parameters $\theta$. $L$ is any single number computed
from the ensemble's solutions; in a calibration, for example,
\begin{equation}
  L(\theta) \;=\; \sum_{k=1}^{n} \ell\bigl(y^{(k)}(t_1;\theta)\bigr),
  \label{eq:loss}
\end{equation}
where $y^{(k)}(t_1;\theta)$ is the final state of the $k$-th trajectory,
which depends on the parameters through the solve, and $\ell$ measures
each trajectory's misfit against its observed data. The parameters enter $L$
only through the solutions, so computing $\nabla_\theta L$ means
differentiating the solve itself.
The gradient has one component for each of the $m$ components of
$\theta$, where $m$ can be a
handful of physical constants or the thousands of weights of a neural
right-hand side. Automatic differentiation computes such derivatives by
applying the chain rule to the operations a program actually executes
\citep{griewank2008evaluating, baydin2018automatic}, in one of two directions.
Forward mode pushes one direction of parameter change through the
computation alongside the ordinary values, yielding the derivative in that
direction after one pass; the full gradient therefore needs $m$
passes, which is affordable when $m$ is small \citep{alexe2009forward}. Reverse mode
runs the program forward once, then sweeps backward from the loss, propagating a sensitivity, called an adjoint, through each operation in
reverse order; one sweep yields the whole gradient, for any $m$. With one scalar loss, reverse mode needs fewer passes than forward mode for any $m$ above one. A backward sweep over an adaptive loop, however, costs several times as much as one forward-mode pass (\cref{sec:benchmarks-forward-mode}), so for a handful of inputs per trajectory forward mode remains the practical choice. For training neural right-hand sides, where $m$ runs to thousands, reverse mode is the only one \citep{chen2018neuralode}. Its price is twofold: the backward sweep needs the
intermediate values of the forward pass, in reverse order, so they must be
stored or recomputed \citep{griewank2000revolve}; and it must revisit every
operation the forward pass executed that could depend on $\theta$.

The adjoint comes in two forms. Applied to the numerical steps a solver
executed, it is a \emph{discrete} adjoint; solving a separate adjoint
differential equation backward in time instead gives a \emph{continuous}
adjoint \citep{chen2018neuralode}, which is not in general the derivative of
the numerical solution returned \citep{zhuang2020aca, gholami2019anode}.

\subsection{Why adaptive stepping and reverse mode conflict}\label{sec:diff-controller}

That last requirement is where the conflict lies: reverse mode differentiates
the program as it actually ran, with every operation the parameters
influenced, rather than the equation the program was written to solve. In an adaptive
solve, the parameters enter the stage evaluations of $f$, and from there the
error estimate, the step-size choice, and every rejected trial also depend on
$\theta$. Solvers usually exclude the step-size choice from the derivative. \diffrax{}, for example, differentiates through the stage evaluations but treats the quantities that only choose the step size (the initial step, the error norm, and the step-size factor) as fixed constants \citep[\S 5.4.2.2]{kidger2022neural}; torchode does the same. The derivative each returns is therefore that of the solve with the accepted step sizes held fixed.
The loop itself remains to be differentiated. Reverse mode has to store or recompute the state of every iteration and then visit the iterations in reverse order. Both the storage and the backward sweep have to be laid out before the number of iterations is known, so the loop is given a fixed upper bound on its length and the backward sweep is arranged for that bound. Storing every intermediate value of a long loop exhausts memory, so
implementations keep only a subset, called checkpoints, and recompute the values between them during the backward sweep
\citep{griewank2000revolve, stumm2010new, wang2009minimal}; this trade of memory for repeated computation is checkpointing. This overhead is what the backward sweep of a differentiated adaptive loop pays; the forward solve itself is not slowed by being differentiable, and its
gap to the fused kernels in \cref{sec:benchmarks-forward} comes from the
lockstep execution described next.

Besides being slow, the differentiated adaptive solve is not well defined at
certain points. The accept/reject test is a branch on a computed number, and a small
change in $\theta$ can move a trial across that threshold, discarding one
downstream step sequence and choosing another. At such a boundary the solve is a discontinuous function of its
parameters, and a derivative accounting for the branching does not exist
there. Any differentiable solver therefore returns the derivative of the computation
with its branch decisions held fixed, that is, of one particular accepted
step sequence, and different choices of what to hold fixed give measurably
different derivatives \citep{ma2021comparison}. Which version
\gradsolve{} chooses, and what its gradient means, is stated in
\cref{sec:method}.

\subsection{Three ways to run an ensemble on a GPU}\label{sec:diff-gpu}

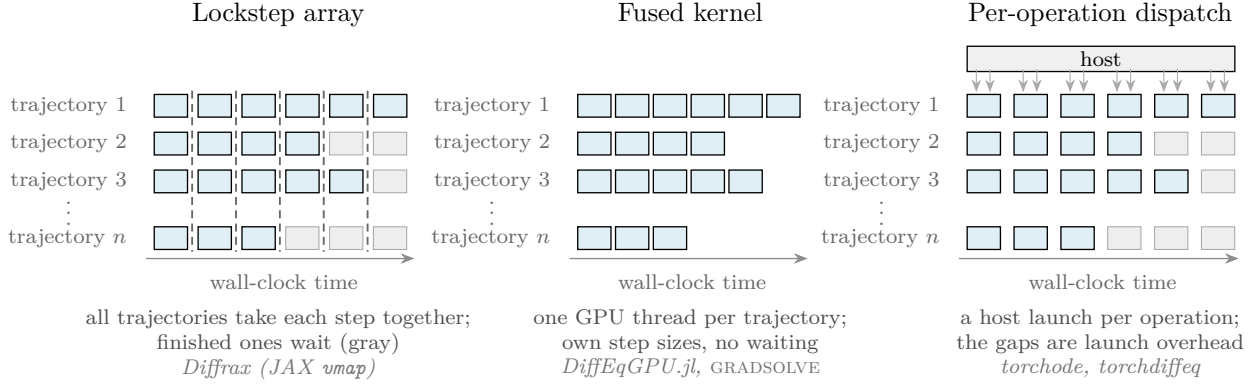
\begin{figure}[tb]
  \centering
  \makebox[\textwidth][c]{\definecolor{emblue}{HTML}{0072B2}%
\definecolor{emghost}{HTML}{AAAAAA}%
\begin{tikzpicture}[
    x=1cm, y=1cm, >=Stealth, font=\footnotesize,
    blk/.style={draw=black, fill=emblue!12, minimum width=0.44cm, minimum height=0.30cm,
                inner sep=0pt, line width=0.5pt, anchor=west},
    idle/.style={draw=emghost, fill=emghost!20, minimum width=0.44cm, minimum height=0.30cm,
                 inner sep=0pt, line width=0.4pt, anchor=west},
    gname/.style={font=\small, anchor=base, align=center},
    gsub/.style={font=\scriptsize, text=black!70, anchor=north, align=center},
    gtool/.style={font=\scriptsize\itshape, text=black!55, anchor=north, align=center},
    axgrey/.style={black!45, line width=0.5pt},
    axlab/.style={anchor=east, font=\scriptsize, text=black!60},
    hostbar/.style={draw=black, fill=black!6, minimum height=0.26cm, inner sep=2pt,
                    line width=0.5pt, font=\scriptsize, anchor=west},
  ]
  \def\yA{0} \def\yB{-0.5} \def\yC{-1.0}   %
  \def\p{0.5}                              %
  \def\yTitle{1.12}                        %
  \def\ySub{-2.55}                         %
  \newcommand{\panelaxes}[1]{%
    \node[axlab] at (-0.22,\yA) {trajectory 1};
    \node[axlab] at (-0.22,\yB) {trajectory 2};
    \node[axlab] (tthree) at (-0.22,\yC) {trajectory 3};
    \node[font=\scriptsize, text=black!60, anchor=north, inner sep=1pt]
        at ([yshift=-8pt]tthree.south) {$\smash{\vdots}$};
    \node[axlab] at (-0.22,-1.75) {trajectory $n$};
    \draw[->, axgrey] (-0.10,-2.02) -- (#1,-2.02);
    \node[anchor=north, font=\scriptsize, text=black!60] at (#1*0.5,-2.10) {wall-clock time};%
  }

  \begin{scope}[shift={(0,0)}]
    \def\pl{0.58}   %
    \panelaxes{3.45}
    \foreach \i in {0,...,5} {\node[blk] at (\i*\pl,\yA) {};}
    \foreach \i in {0,...,3} {\node[blk] at (\i*\pl,\yB) {};}
    \foreach \i in {4,5}     {\node[idle] at (\i*\pl,\yB) {};}
    \foreach \i in {0,...,4} {\node[blk] at (\i*\pl,\yC) {};}
    \foreach \i in {5}       {\node[idle] at (\i*\pl,\yC) {};}
    \foreach \i in {0,...,2} {\node[blk] at (\i*\pl,-1.75) {};}
    \foreach \i in {3,...,5} {\node[idle] at (\i*\pl,-1.75) {};}
    \foreach \i in {1,...,5} {\draw[densely dashed, black!60, line width=0.7pt]
        (\i*\pl-0.07,0.20) -- (\i*\pl-0.07,-1.93);}
    \node[gname] at (1.65,\yTitle) {Lockstep array};
    \node[gsub]  at (1.65,\ySub) {all trajectories take each step together;\\ finished ones wait (gray)};
    \node[gtool] at (1.65,-3.20) {Diffrax (JAX \texttt{vmap})};
  \end{scope}

  \begin{scope}[shift={(5.6,0)}]
    \panelaxes{3.05}
    \foreach \i in {0,...,5} {\node[blk] at (\i*\p,\yA) {};}   %
    \foreach \i in {0,...,3} {\node[blk] at (\i*\p,\yB) {};}   %
    \foreach \i in {0,...,4} {\node[blk] at (\i*\p,\yC) {};}   %
    \foreach \i in {0,...,2} {\node[blk] at (\i*\p,-1.75) {};}  %
    \node[gname] at (1.5,\yTitle) {Fused kernel};
    \node[gsub]  at (1.5,\ySub) {one GPU thread per trajectory;\\ own step sizes, no waiting};
    \node[gtool] at (1.5,-3.20) {DiffEqGPU.jl, \gradsolve{}};
  \end{scope}

  \begin{scope}[shift={(10.75,0)}]
    \def\q{0.62}                           %
    \panelaxes{3.60}
    \node[hostbar, minimum width=3.54cm] at (0,0.62) {host};
    \foreach \i in {0,...,5} {\draw[->, emghost, line width=0.4pt]
        (\i*\q+0.13,0.48) -- (\i*\q+0.13,0.18);
      \draw[->, emghost, line width=0.4pt]
        (\i*\q+0.31,0.48) -- (\i*\q+0.31,0.18);}
    \foreach \i in {0,...,5} {\node[blk] at (\i*\q,\yA) {};}
    \foreach \i in {0,...,3} {\node[blk] at (\i*\q,\yB) {};}
    \foreach \i in {4,5}     {\node[idle] at (\i*\q,\yB) {};}
    \foreach \i in {0,...,4} {\node[blk] at (\i*\q,\yC) {};}
    \foreach \i in {5}       {\node[idle] at (\i*\q,\yC) {};}
    \foreach \i in {0,...,2} {\node[blk] at (\i*\q,-1.75) {};}
    \foreach \i in {3,...,5} {\node[idle] at (\i*\q,-1.75) {};}
    \node[gname] at (1.77,\yTitle) {Per-operation dispatch};
    \node[gsub]  at (1.77,\ySub) {a host launch per operation;\\ the gaps are launch overhead};
    \node[gtool] at (1.77,-3.20) {torchode, torchdiffeq};
  \end{scope}
\end{tikzpicture}}
  \caption{The same ensemble of $n$ trajectories under the three GPU
  execution designs, showing the first three trajectories, needing six,
  four, and five accepted steps, and the last, the $n$-th, needing three. Each
  row is one trajectory. Each block is one
  accepted step of the adaptive loop of \cref{sec:diff-adaptive}: the
  Runge--Kutta stage evaluations of the right-hand side of the ODE, followed
  by the error test and the controller decision. Block width is uniform and does not represent step size, and rejected trials are not drawn. Horizontal position tracks wall-clock time in every panel, along the bottom arrow; the white gaps between steps in the right panel are the launch overhead, drawn as blank space. Left: a lockstep
  array program advances all trajectories together, one loop iteration per
  dashed line, in which every trajectory takes one trial step of its own
  size, so trajectories that finish early wait through the remaining
  iterations (gray). Center: a fused kernel gives each trajectory its
  own thread, with its own step sizes, so each row ends when its work
  ends.
  Right: per-operation dispatch launches each operation of each step (stage
  evaluations, error norm, controller update) as its own kernel from the
  host (the CPU that drives the GPU); two launch arrows are drawn per step; finished trajectories wait (gray) as in
  the left panel, and the white gaps between steps are the launch overhead,
  stretching the same work over more wall-clock time. \gradsolve{} builds on
  the fused design.}
  \label{fig:execmodels}
\end{figure}

The costs above concern one trajectory. Running thousands on a GPU adds two hardware facts that shape every design.
First, a GPU executes threads in fixed-size groups that issue the same instruction together, so when threads in a
group branch differently, the group runs both branches one after the other
\citep{lindholm2008nvidia}. Second, launching a kernel from the host (the CPU that drives the GPU) carries a fixed
overhead, however little work the kernel does.
Three designs are in established use. They place the adaptive loop on the
hardware in different ways, inherit different costs from those two facts, and
are sketched, left to right, in \cref{fig:execmodels}.

\begin{description}
\item[Lockstep array (left panel of \cref{fig:execmodels}).] This design maps a single-trajectory
solver over the whole ensemble with an array transformation, such as
\code{vmap} in JAX \citep{bradbury2018jax}; \diffrax{} \citep{kidger2021diffrax} under \code{vmap} is the mature example;
the lockstep is a property of the array transformation, not of the solver. Each iteration of the
loop advances the whole ensemble by one trial step, trajectories whose trial was rejected keep their old state and retry in the
next iteration, and trajectories that have already finished are carried along
unchanged; neither is skipped. The solve is then an ordinary array program, so
the framework's reverse-mode differentiation applies to it, with the bounded
loop and checkpointing of \cref{sec:diff-controller}; the further cost is that the whole ensemble keeps stepping until its slowest trajectory finishes, and the backward sweep likewise carries every trajectory through every iteration.

\item[Fused kernel (center panel of \cref{fig:execmodels}).] This design compiles the entire
adaptive loop, including its branches, into a single kernel in which each
GPU thread advances its own trajectory, keeping its state in registers,
the small, fast memory private to each thread; DiffEqGPU.jl
\citep{utkarsh2024diffeqgpu, rackauckas2017differentialequations} is the
published instance. Nothing is launched per step, and no thread waits for
a slow neighbor beyond the group-level serialization above, which makes
this the fastest way to run an ensemble forward. Its limits are state
size and the lack of reverse-mode differentiation. A thread can use at most 255 registers, each holding one 32-bit word
\citep{nvidia2025cuda}, so beyond a modest state dimension the state and stage
vectors of a trajectory no longer fit and the compiler moves them to slower
memory. The in-kernel branching is exactly the loop that \cref{sec:diff-controller}
showed reverse mode cannot pass through cheaply. DiffEqGPU.jl differentiates inside these kernels in forward mode only (\cref{sec:benchmarks-reverse}).

\item[Per-operation dispatch (right panel of \cref{fig:execmodels}).] This design keeps the loop
on the host and launches each solver operation (stage evaluation, error
norm, controller update) as its own kernel; the PyTorch solvers
torchode \citep{lienen2022torchode} and torchdiffeq
\citep{chen2018torchdiffeq} work this way. In torchode each trajectory
carries its own step-size controller, so no trajectory is forced onto
another's step sizes, although the batch loop still runs until its slowest
trajectory finishes, with finished trajectories carried along as in the
lockstep design; and because every operation is an ordinary framework
operation, reverse mode applies directly. torchdiffeq instead shares one
adaptive step sequence across the batch, so it waits for the slowest
trajectory, and the interface we benchmark, \code{odeint\_adjoint}, solves a
continuous adjoint equation backward in time rather than differentiating the operations themselves
(\cref{sec:diff-admodes}); its plain \code{odeint} interface
differentiates the operations directly. The cost is
the fixed launch overhead, paid several times per step of the integration.
\end{description}

Each design thus favors one of the two requirements, forward speed and
reverse-mode differentiability, over the other: the fused kernels
give the highest forward speed, while those that support reverse-mode
differentiation pay for it in lockstep stepping or per-operation dispatch.
All of that work (the error tests, the controller, the rejected trials) serves
only to find step sizes that are not known in advance; once a solve has
run, they are known, and \cref{sec:method} builds \gradsolve{} on that
observation.

\section{The \gradsolve{} method: record and replay}\label{sec:method}

\gradsolve{} differentiates a second, simpler computation built from the
adaptive solve, rather than the adaptive solve itself. This section first describes that
construction (\cref{sec:method-construction}), then the integration methods it
is built around (\cref{sec:method-integrators}), then the implementations of
it inside the library and how one is chosen for a request
(\cref{sec:method-engines}), and finally the two calls a user sees
(\cref{sec:method-interface}).

\subsection{The construction}\label{sec:method-construction}

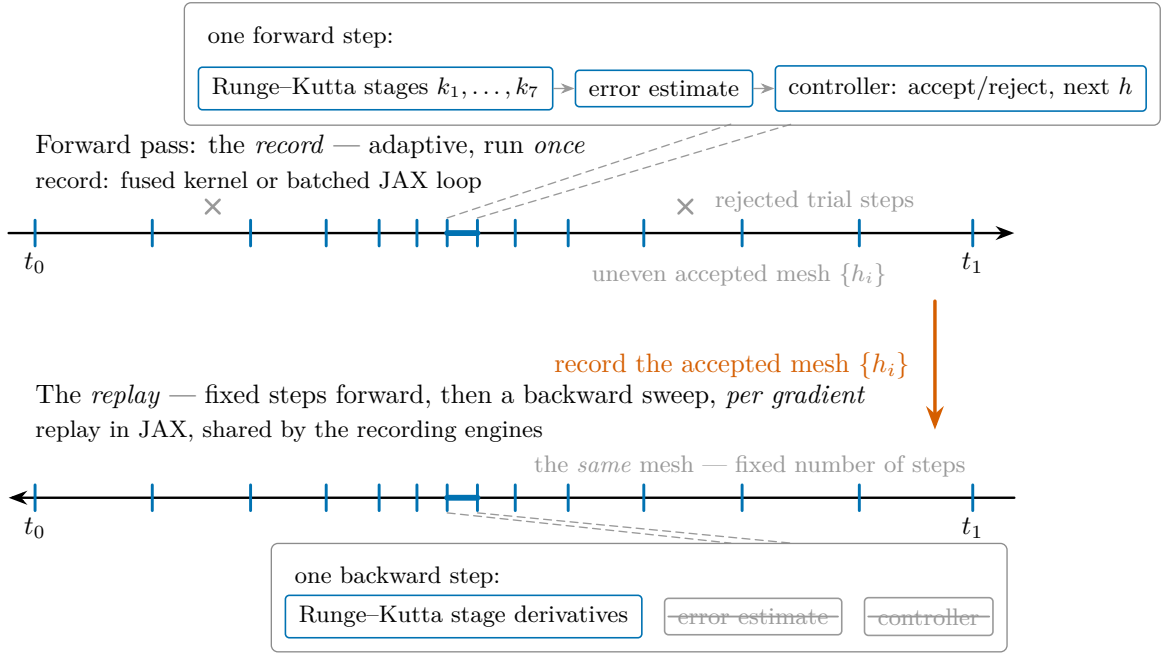
\begin{figure}[tb]
  \centering
  \definecolor{mechblue}{HTML}{0072B2}%
\definecolor{mechverm}{HTML}{D55E00}%
\definecolor{mechghost}{HTML}{999999}%
\begin{tikzpicture}[
    x=1cm, y=1cm, >=Stealth, line cap=round,
    lane/.style={->, line width=0.9pt, draw=black},
    tickmark/.style={draw=mechblue, line width=1.1pt},
    rejected/.style={draw=mechghost, line width=1.0pt},
    zoom/.style={draw=mechghost, densely dashed, line width=0.5pt},
    stagebox/.style={draw=mechblue, line width=0.6pt, rounded corners=1.5pt,
                     inner xsep=5pt, inner ysep=4pt, font=\footnotesize, fill=white},
    stageghost/.style={stagebox, draw=mechghost, text=mechghost},
    lanelabel/.style={font=\small, anchor=west, inner sep=0pt},
    stepgroup/.style={draw=black!45, line width=0.5pt, rounded corners=2.5pt,
                      inner sep=5pt},
    sublabel/.style={font=\footnotesize, text=mechghost, inner sep=1pt},
  ]

  \def\meshticks{0.0, 1.55, 2.85, 3.85, 4.55, 5.05, 5.45, 5.85, 6.35, 7.05, 8.05, 9.35, 10.9, 12.4}
  \def\lanelen{12.4}

  \begin{scope}[shift={(1.5,4.55)}]
    \node[lanelabel, align=left] at (0,0.92)
      {Forward pass: the \emph{record} --- adaptive, run \emph{once}\\[1pt]
       {\footnotesize record: fused kernel or batched JAX loop}};
    \draw[lane] (-0.35,0) -- (\lanelen+0.55,0);
    \node[font=\small, below=3pt] at (0,-0.02) {$t_0$};
    \node[font=\small, below=3pt] at (\lanelen,-0.02) {$t_1$};
    \foreach \xi in \meshticks {\draw[tickmark] (\xi,-0.17) -- (\xi,0.17);}
    \foreach \xr in {2.35, 8.6} {
      \draw[rejected] (\xr-0.09,0.26) -- (\xr+0.09,0.44);
      \draw[rejected] (\xr-0.09,0.44) -- (\xr+0.09,0.26);
    }
    \node[sublabel, anchor=west] at (8.95,0.42) {rejected trial steps};
    \node[sublabel, anchor=east] at (11.3,-0.55) {uneven accepted mesh $\{h_i\}$};
    \draw[mechblue, line width=2.2pt] (5.45,0) -- (5.85,0);
    \draw[zoom] (5.45,0.20) -- (9.2,1.44);
    \draw[zoom] (5.85,0.20) -- (10.0,1.44);
    \node[stagebox, anchor=south west] (fstages) at (2.15,1.62)
      {Runge--Kutta stages $k_1,\dots,k_7$};
    \node[stagebox, anchor=west] (ferr) at ($(fstages.east)+(0.28,0)$)
      {error estimate};
    \node[stagebox, anchor=west] (fctrl) at ($(ferr.east)+(0.28,0)$)
      {controller: accept/reject, next $h$};
    \draw[->, mechghost, line width=0.5pt] (fstages.east) -- (ferr.west);
    \draw[->, mechghost, line width=0.5pt] (ferr.east) -- (fctrl.west);
    \node[font=\footnotesize, anchor=south west] (flab) at (2.15,2.32) {one forward step:};
    \node[stepgroup, fit=(flab)(fstages)(ferr)(fctrl)] {};
  \end{scope}

  \draw[->, mechverm, line width=1.3pt] (13.4,3.65) -- (13.4,1.95);
  \node[font=\small, text=mechverm, anchor=east] at (13.2,2.8)
    {record the accepted mesh $\{h_i\}$};

  \begin{scope}[shift={(1.5,1.05)}]
    \node[lanelabel, align=left] at (0,1.12)
      {The \emph{replay} --- fixed steps forward, then a backward sweep, \emph{per gradient}\\[1pt]
       {\footnotesize replay in JAX, shared by the recording engines}};
    \draw[lane] (\lanelen+0.55,0) -- (-0.35,0);   %
    \node[font=\small, below=3pt] at (0,-0.02) {$t_0$};
    \node[font=\small, below=3pt] at (\lanelen,-0.02) {$t_1$};
    \foreach \xi in \meshticks {\draw[tickmark] (\xi,-0.17) -- (\xi,0.17);}
    \node[sublabel, anchor=east] at (12.35,0.42)
      {the \emph{same} mesh --- fixed number of steps};
    \draw[mechblue, line width=2.2pt] (5.45,0) -- (5.85,0);
    \draw[zoom] (5.45,-0.20) -- (9.2,-0.60);
    \draw[zoom] (5.85,-0.20) -- (10.0,-0.60);
    \node[stagebox, anchor=north west] (bstages) at (3.3,-1.28)
      {Runge--Kutta stage derivatives};
    \node[stageghost, anchor=west] (berr) at ($(bstages.east)+(0.28,0)$)
      {error estimate};
    \node[stageghost, anchor=west] (bctrl) at ($(berr.east)+(0.28,0)$)
      {controller};
    \draw[mechghost, line width=0.7pt] ($(berr.west)+(0.06,0)$) -- ($(berr.east)+(-0.06,0)$);
    \draw[mechghost, line width=0.7pt] ($(bctrl.west)+(0.06,0)$) -- ($(bctrl.east)+(-0.06,0)$);
    \node[font=\footnotesize, anchor=north west] (blab) at (3.3,-0.78) {one backward step:};
    \node[stepgroup, fit=(blab)(bstages)(berr)(bctrl)] {};
  \end{scope}

\end{tikzpicture}
  \caption{Record and replay. One adaptive forward solve runs per trajectory
  and records the accepted step sizes $\{h_i\}$, which fix the mesh. The magnified boxes show
  the anatomy of one step: the seven Runge--Kutta stages
  $k_1, \dots, k_7$ of the Tsit5 pair, followed by the error estimate and the
  controller decision. Gradients are computed by differentiating a fixed-step
  replay of the recorded steps. The mesh may be recorded either by a fused GPU
  kernel or by a batched loop written in JAX, and the recording engines then
  differentiate the same JAX replay of it. In the replay the error estimate and the
  controller are struck out because the steps are already decided; they and
  the rejected trials are absent from the differentiated computation, which is
  therefore a fixed-length chain with no branch in it (\cref{sec:whyfast}). The arrow in the lower part of the figure runs from $t_1$ back to $t_0$
  because it depicts the backward sweep; the fixed-step replay that precedes
  it walks the same mesh forward.}
  \label{fig:recordreplay}
\end{figure}

\Cref{fig:recordreplay} sketches it. A first pass, the \emph{record}, runs an
ordinary adaptive solver exactly as it would run if no gradient were wanted,
and keeps one extra piece of information: the sequence of step sizes it
accepted. These fix the time points of the solve, its \emph{mesh}. Rejected
trials leave no trace in the record. A second pass, the \emph{replay}, walks
the same right-hand side through the same steps, with every step size now a
fixed number handed in as data. The replay uses the recorded step sizes
directly, so it needs no error estimate, makes no step-size choice and takes
no branch; this is the computation \gradsolve{} differentiates.

Across an ensemble the records differ in length, because each trajectory
accepts its own number of steps (\cref{sec:diff-adaptive}). \gradsolve{}
therefore pads every record with zeros to the length of the longest one. A
step of size zero leaves the state unchanged, so the padded rows contribute
nothing to the result or to its derivative. The padded records form one
rectangular array, and the whole ensemble replays as a single regular
computation, which GPUs execute efficiently.

The replay of one trajectory is a chain of $S$ one-step maps,
\begin{equation}
  y_i = \Phi_i(y_{i-1};\, h_i, \theta), \qquad i = 1, \dots, S,
  \label{eq:replay}
\end{equation}
where $\Phi_i$ is one step of the underlying method and $S$ is the recorded
step count. For a non-stiff method, $\Phi_i$ is the explicit map of
\cref{eq:stages,eq:update} with the recorded $h_i$ in place of $h$; a stiff
method uses a different one-step map, described in \cref{sec:method-integrators}. Because these are the very
steps the adaptive solver accepted, running the chain reproduces the
trajectory the solver already produced. Because the chain has no
branching left in it, reverse-mode differentiation of it is the textbook
case \citep{griewank2008evaluating}: one backward sweep through the $S$ maps
returns the exact gradient of whatever the chain computes.

That statement is the precise sense in which \gradsolve{}'s gradient is
exact. Two maps make it formal. The \emph{record} map $R$ runs the adaptive
solver at the base parameters $\theta_0$ and returns the sequence of accepted
step sizes, $h = R(\theta_0)$; the \emph{replay} map $\Psi$ composes the chain
of \cref{eq:replay} with that mesh held fixed as data,
$y_S = \Psi\bigl(\theta;\, \operatorname{sg}[h]\bigr)$, where
$\operatorname{sg}[\cdot]$ is the stop-gradient that makes $h$ a constant of the
differentiation. \gradsolve{} returns $\partial \Psi/\partial \theta$ at
$\theta_0$: the exact reverse-mode derivative of the replay, with the recorded
step sizes held fixed as data. Only the step sizes survive the record: the replay
recomputes the forward states from them, and the adaptive pass's own intermediate values are
discarded along with its rejected trials. The alternative --- retaining that forward pass and
excising the rejected steps from its backward sweep alone --- would carry each trajectory's
accept/reject structure into the differentiated program, which is precisely what the fixed-length
chain removes; the recomputed forward pass is a minority of the gradient's cost, the backward
sweep dominating (\cref{sec:whyfast}). When one record is reused across an optimization, the derivative is taken at the current parameters $\theta$ while the mesh stays $h = R(\theta_0)$, so the returned quantity is $\partial \Psi\bigl(\theta;\, \operatorname{sg}[R(\theta_0)]\bigr)/\partial \theta$; \cref{sec:benchmarks-fitting} measures how far this departs from a fresh record as $\theta$ moves. Because $\operatorname{sg}$ severs the dependence
of $h$ on $\theta$, this derivative drops the term
$(\partial \Psi/\partial h)\,(\partial R/\partial \theta)$ by which the mesh
would move with the parameters. It accounts for how
the parameters, and the initial state, enter the solve at every accepted
step; it does not account for how an adaptive solver would have chosen
different steps at nearby values of them. How much this matters over a
fitting run --- an optimization loop that repeatedly adjusts the parameters
to fit data, as in a calibration --- is measured in \cref{sec:benchmarks-fitting}: the gradient
direction on the recorded mesh stays within hundredths of a degree of one
from a fresh record. This is the same convention as \diffrax{}'s checkpointed
adjoint, which also stops the gradient at its step-size choices
(\cref{sec:diff-controller}) and returns the discrete adjoint of its own
accepted steps, dropping the same mesh-shift term; on the same method and the
same mesh $R(\theta_0)$ the two derivatives coincide. That the replay's gradient
is not only self-consistent but an accurate sensitivity --- it matches an
independent high-accuracy finite-difference reference to the method's own
forward error --- is shown in \ref{app:gradchecks}. The correctness checks in this paper respect the convention too:
the finite-difference comparisons perturb the same inputs inside the same
recorded replay, so both sides differentiate the same computation. Two
neighboring constructions differ from it: the adaptive checkpoint adjoint of
\citet{zhuang2020aca} also backpropagates only through the accepted steps,
one trajectory at a time in PyTorch, advancing each solver operation as its
own dispatched call. \gradsolve{} recasts that idea as a single zero-padded
fixed-length replay over the whole ensemble, compiled as one GPU program --- which
is what the cost measurements of \cref{sec:whyfast} turn on; run per operation on
the CPU host, as in that work, the same construction would sit in the dispatch-bound
class of the PyTorch solvers of \cref{sec:benchmarks-reverse}, and reversible
integrators \citep{zhuang2021mali} reconstruct the forward states instead of storing
them. The library-level discrete-adjoint solvers --- FATODE
\citep{zhang2014fatode}, PETSc TSAdjoint \citep{zhang2022tsadjoint}, and CVODES in
SUNDIALS \citep{serban2005cvodes, hindmarsh2005sundials} --- return this same class
of derivative on CPU and array-based paths; \gradsolve{} returns it for a fused
per-trajectory GPU ensemble. Within the replay the same choice is a tuning option: it can recompute
its intermediate values during the backward sweep rather than store them
(the library's \code{remat} option, on by default for non-stiff problems and
for stiff systems of 16 or more variables), trading time for memory.

\subsection{The integrators}\label{sec:method-integrators}

The record pass uses the same methods as any adaptive solve, and every method
supplies the same two ingredients: a trial step with an error estimate for
the record, and a plain step for the replay. For non-stiff problems the
default is the Tsit5 pair of \citet{tsitouras2011runge}, with the Verner
seventh-order pair \citep{verner2010numerically} available on request. For
stiff problems the default is the fifth-order Rosenbrock--Wanner method
Rodas5P \citep{steinebach2023rodas5p}; the right-hand sides available in
the fused kernels' own form use instead the second-order linearly implicit
Rosenbrock23 method \citep{rosenbrock1963general, shampine1997matlab}, with
Rodas5P again available on request. Rodas5P carries a continuous extension,
an interpolant inside each step, so that, when it is requested, it can return the state at
intermediate times without taking extra steps; the default for every replay engine instead
integrates one extra step to each requested time (\cref{sec:method-interface}). Rodas5P also
carries the time derivative $\partial f/\partial t$; Rosenbrock23 assumes an autonomous system, one in
which $f$ has no explicit dependence on $t$, and it serves only the stiff
systems with a fused kernel, which are all autonomous. A stiff problem with
explicit time dependence therefore takes Rodas5P without a special request.

A stiff problem changes the one-step map $\Phi_i$ but not the construction
around it. Each Rosenbrock stage solves a linear system with the matrix
$I - \gamma h J$, where $J = \partial f/\partial y$ and $\gamma$ is a method
coefficient, and the recorded mesh wraps around those stages exactly as
before. Differentiating a linear solve does not require differentiating the
steps of Gaussian elimination: if $x$ solves $Ax = b$, the backward sweep
needs only one more solve, against $A^{\top}$ \citep{giles2000adjoint}. JAX
applies this rule to its linear solve automatically, so the stiff replay
inherits it without any special code. The user does not supply $J$ on the
JAX path either: it comes from automatic differentiation of the right-hand side, so
the backward sweep differentiates through it as well as through the stages.

\subsection{The engines and the router}\label{sec:method-engines}

\begin{figure}[tb]
  \centering
  \includegraphics[width=\textwidth]{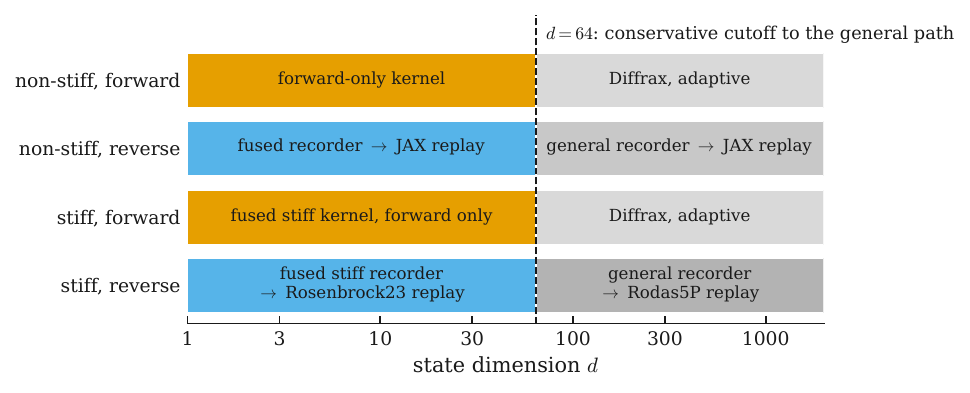}
  \caption{Engine dispatch. Each row is one kind of request (non-stiff or
  stiff, forward solve or gradient) and the horizontal axis is the state
  dimension $d$; the boxes name the engine of \cref{tab:engines} the router
  selects. Gold marks the two fused forward-only kernels, blue the
  fused recorder followed by the differentiated replay, and gray the general
  path, taken by every right-hand side without a fused kernel and by all
  states beyond $d = 64$.}
  \label{fig:routing}
\end{figure}

Each implementation of a solve or gradient route inside the library is an
\emph{engine}. \Cref{tab:engines} lists them by the names used in this paper
and in the code, and \cref{fig:routing} shows which one the router selects
for a request. A \emph{kernel} here is one program launched on the GPU;
a \emph{fused} kernel runs a whole integration inside that one program, one
GPU thread per trajectory, so the fixed cost of a launch is paid once per
ensemble rather than once per step. A fused kernel needs the right-hand side
in its own form, and a \emph{registered field} is a right-hand side available
in that form. The library provides it hand-written for the Lorenz, Van der
Pol, Lorenz-96 and linear test systems and for the stiff Robertson, HIRES and
linear stiff test systems; for any other right-hand side the user registers
the JAX function, and the library translates it (\cref{sec:method-interface}).

\begin{figure}[tb]
  \centering
  \definecolor{dfgold}{HTML}{E69F00}%
\definecolor{dfblue}{HTML}{56B4E9}%
\definecolor{dfgrey}{HTML}{C8C8C8}%
\resizebox{\textwidth}{!}{%
\begin{tikzpicture}[
    x=1cm, y=1cm, >=Stealth, line cap=round,
    box/.style={draw=black!55, line width=0.6pt, rounded corners=3pt,
                align=center, font=\footnotesize, inner sep=4pt,
                minimum height=0.95cm},
    data/.style={box, fill=white},
    arr/.style={->, line width=0.8pt, draw=black!75},
  ]

  \node[data, text width=2.3cm] (rhs) at (0,0.1)
    {right-hand side, a JAX function written by the user};

  \node[box, fill=dfgold, text width=2.7cm] (fwd) at (3.9,2.85)
    {forward-only kernel (no record)};
  \node[data, text width=1.9cm] (final) at (7.7,2.85) {final states};

  \node[box, fill=dfblue, text width=2.7cm] (frec) at (3.9,1.05)
    {fused recorder or fused stiff recorder (Warp kernels; registered fields)};
  \node[box, fill=dfgrey, text width=2.6cm] (grec) at (3.9,-0.95)
    {general recorder (batched JAX loop; any right-hand side)};
  \node[data, text width=2.3cm] (mesh) at (7.7,0.05)
    {recorded mesh: the accepted step sizes};
  \node[box, fill=dfblue, text width=2.5cm] (replay) at (11.4,0.05)
    {fixed-step replay in JAX, shared by the recording engines};
  \node[data, text width=2.2cm] (grad) at (15.0,0.05)
    {gradient by reverse mode};

  \node[box, fill=dfblue, text width=11.2cm] (adj) at (7.2,-2.75)
    {in-kernel adjoint: record, replay and backward sweep inside Warp kernels
     (registered stiff fields)};

  \draw[arr] (rhs.east) to[out=25,in=180] (fwd.west);
  \draw[arr] (rhs.east) to[out=-25,in=180] (grec.west);
  \draw[arr] (rhs.south) to[out=-90,in=180] (adj.west);
  \draw[arr, densely dashed] (rhs.east) to[out=8,in=180] (frec.west);

  \draw[arr, draw=dfgold!85!black] (fwd.east) -- (final.west);

  \draw[arr] (frec.east) to[out=0,in=150] (mesh.north west);
  \draw[arr] (grec.east) to[out=0,in=210] (mesh.south west);
  \draw[arr] (mesh.east) -- (replay.west);
  \draw[arr] (replay.east) -- (grad.west);

  \draw[arr, draw=dfblue!70!black] (adj.east) to[out=0,in=-90] (grad.south);

\end{tikzpicture}%
}
  \caption{The data path of one request. A user right-hand side enters at the
  left and takes one of three routes. On the forward-only route (gold) a single
  fused kernel integrates to the final states and records nothing. On the
  gradient route the mesh, the sequence of accepted step sizes, is recorded
  either by the fused recorder (a Warp kernel, for a right-hand side in the
  kernels' own form, built in or translated from the user's JAX function) or by
  the general recorder (a batched JAX loop, for any right-hand side as
  written); the two
  recorders produce the same mesh and feed one fixed-step replay written in JAX
  (blue), whose reverse mode returns the gradient. The dashed arrow marks the
  translation of \cref{sec:method-interface}, by which a user function becomes a
  registered field the fused recorder can call. The in-kernel adjoint (blue,
  bottom) is the one route that keeps the record, the replay and the backward
  sweep inside the Warp kernels, for the registered stiff fields, and delivers
  the same gradient. Gold marks the
  forward-only kernel, blue the fused recorder and replay, and gray the general
  path.}
  \label{fig:dataflow}
\end{figure}
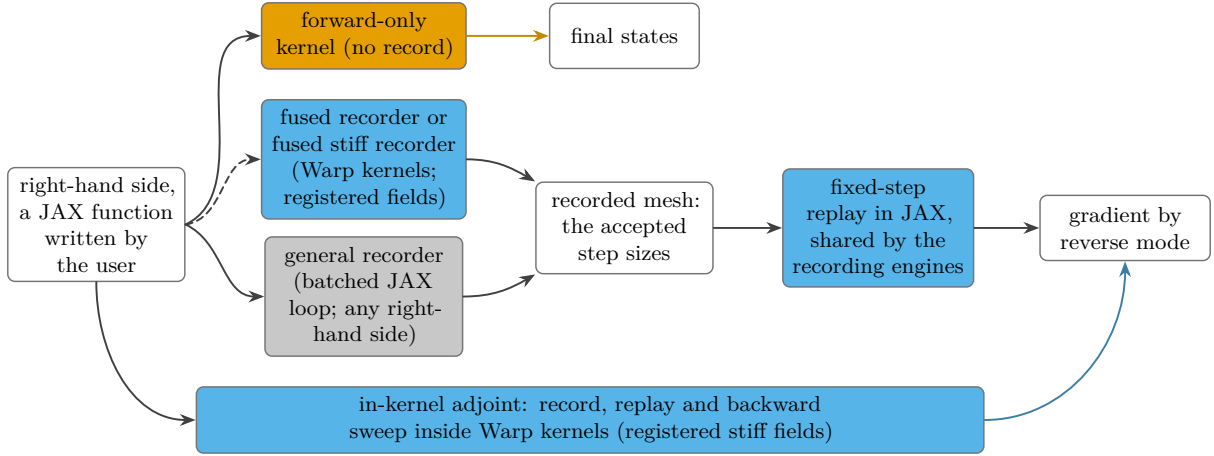

\begin{table}[!tb]
  \centering\footnotesize
  \caption{The engines of \gradsolve{}. Each row is one implementation of a
  solve or gradient route. The second column gives the name reported on the
  result and accepted by the \code{engine} argument. A registered field is a
  right-hand side available in the fused kernels' own form, hand-written for
  the built-in systems or translated from the user's JAX function. The fused
  Warp kernels \code{warp\_ode} and \code{warp\_rosenbrock} each name a single
  kernel that the router runs without a record for a forward-only solve and
  with the record kept when a gradient is wanted; the gradient route of
  \code{warp\_ode} is reported as \code{warp\_replay}, so only
  \code{warp\_rosenbrock} labels two rows. The
  last column states the router's default choice; $d$ is the state
  dimension.}
  \label{tab:engines}
  \setlength{\tabcolsep}{4pt}
  \begin{tabular}{@{}>{\raggedright\arraybackslash}p{0.21\textwidth}>{\raggedright\arraybackslash}p{0.19\textwidth}>{\raggedright\arraybackslash}p{0.27\textwidth}>{\raggedright\arraybackslash}p{0.27\textwidth}@{}}
    \toprule
    Engine & In the code & Record and replay & Default for \\
    \midrule
    Forward-only kernel & \code{cuda\_tsit5}, \code{warp\_ode} & no record; one fused kernel solves to the final state & registered non-stiff field, no gradient wanted, $d \le 64$ \\
    \addlinespace
    Fused stiff kernel, forward only & \code{warp\_rosenbrock} & no record; one fused kernel & registered stiff field, no gradient wanted, $d \le 64$ \\
    \addlinespace
    Fused recorder $\to$ JAX replay & \code{warp\_replay} & record in a fused kernel; Tsit5 replay in JAX & registered non-stiff field, gradient wanted, $d \le 64$ \\
    \addlinespace
    Fused stiff recorder $\to$ Rosenbrock23 replay & \code{warp\_rosenbrock} & record in a fused kernel; Rosenbrock23 replay in JAX & registered stiff field, gradient wanted, $d \le 64$ \\
    \addlinespace
    General recorder $\to$ JAX replay (the general path) & \code{tsit5\_replay}, \code{rodas5p\_replay}; \code{vern7\_replay} by name & record as a batched JAX loop on the device; Tsit5 or Rodas5P replay in JAX & every other gradient request: an unregistered field, or $d > 64$ \\
    \addlinespace
    In-kernel adjoint & \code{fused\_rosenbrock\_}\newline\code{backward} & record and replay both in fused kernels & registered stiff field, by name, when memory is the limit \\
    \addlinespace
    \diffrax{} & \code{diffrax} & adaptive solve, differentiated by its own checkpointed adjoint & unregistered field, no gradient wanted; by name as the reverse-mode baseline \\
    \bottomrule
  \end{tabular}
\end{table}

On every gradient route except the in-kernel adjoint (\cref{fig:dataflow}), the replay is the same
computation: a fixed-step loop over the recorded step sizes, written in JAX
\citep{bradbury2018jax} and differentiated by its reverse mode. The engines
differ only in how the mesh is recorded. The \emph{general recorder} runs the
adaptive solve as a batched JAX loop on the device, for any right-hand side,
and places no bound on the state dimension. The \emph{fused recorder} runs the
same adaptive solve as a single fused kernel, written with the Warp kernel
framework \citep{nvidia2024warp}, and runs the registered fields, built in or
translated from the user's JAX function.
The two record the same mesh up to floating-point roundoff (the same accepted
steps, with step sizes agreeing to about $10^{-8}$ relative) and hand it to
the same replay, so they return the same gradient. Since the record is paid
once and the per-gradient time counts only the replay, they also reach nearly
the same per-gradient time (\cref{sec:whyfast}). With either recorder the
record costs about one forward solve of the ensemble on the device
(\cref{sec:benchmarks-highorder} reports the times), so it can be renewed at
any time; \cref{sec:benchmarks-fitting} states how the fitting loops use it.

The \emph{in-kernel adjoint} is the one engine whose replay runs in Warp
rather than in JAX, for the registered stiff fields: a fused kernel steps
forward over the recorded mesh, and Warp's automatic differentiation
generates the matching backward kernel. Its purpose is memory. The backward
kernel does not store the intermediate values of every stage, as reverse mode
normally does; it keeps only the accepted states, one per step, and
recomputes the stage values from them. At a batch of \num{128} stiff
trajectories with a neural right-hand side (\cref{sec:benchmarks-fitting})
its record occupies \memRecordVsTape{}. The replay it is compared against stores every intermediate value; recompute (\code{remat}) is off at this state dimension (\cref{sec:method-construction}), so this is the gap against a store-everything replay rather than a recompute-enabled one. Its advantage over the JAX replay is
memory rather than speed, so it is selected by name; \ref{app:gradchecks}
describes the two parts of it that are written by hand.

The router reads three properties of a request, stiff or not, the state
dimension, and whether a gradient is wanted, and applies the last column of
\cref{tab:engines}. The cutoff $d \le 64$ for the fused engines is
conservative: it is set below the dimension at which a trajectory's state
stops fitting in a thread's registers (\cref{sec:benchmarks-dim}). Every
other request takes the general path, so every request the library accepts
can be differentiated. \diffrax{} serves the forward-only requests the fused
kernels do not cover, when it is installed, and remains available by name as
the reverse-mode baseline, differentiating through an adaptive solve rather
than a recorded replay (\ref{app:protocol-baseline}).

\subsection{The interface}\label{sec:method-interface}

A user needs none of the above to use the library: \gradsolve{} offers one
call to solve an ensemble and one to differentiate it, condensed in
\cref{lst:interface} from the first example in the repository, and the router
chooses the engine.
A user who wants a particular engine can also name it, with the names of
\cref{tab:engines}, through the \code{engine} argument, and the engine that
ran is reported on the result either way. A user-defined problem is any
object with six members. By default an arbitrary user right-hand side takes the general
path: it is differentiated directly through \code{f\_jax}, and its record
runs on the device as a batched loop over the ensemble
(\cref{sec:benchmarks-highorder} reports its time). With \code{fused=True},
or by registering the right-hand side under a name, the library translates
the JAX function into the form the fused kernels call, together with its
Jacobian for a stiff problem, derived by automatic differentiation, and the
same fused engines then run it with no kernel written by hand. The
translation covers right-hand sides built from elementary arithmetic, the
standard transcendental functions and fixed-index array operations; a
right-hand side with data-dependent control flow is refused with a message
naming the operation, and the request takes the general path. On four of the
built-in systems (Lorenz, Van der Pol, Robertson and HIRES) the translated
fields and Jacobians agree with the hand-written kernels to $10^{-15}$ and record the same accepted steps, and on
an A100 a translated Lorenz field returned final states and gradients
identical to the hand-written field's. The returned gradient function differentiates with
respect to the parameters, the initial state, or both. Every JAX replay
engine can also return the state at requested intermediate times: the replay
takes one extra solver step from the preceding mesh point exactly to each
requested time, so the returned state is integrated, not interpolated. For
the registered fields the record and the replay can both run in single
precision, with GPU-verified gradients (\cref{sec:benchmarks-reverse} measures
the saving).

\begin{center}
\begin{minipage}{\linewidth}
\footnotesize
\begin{verbatim}
class Lorenz:                              # a problem: any object with these six members
    name, dim = "lorenz", 3                # a label, and the number of state components
    t0, t1, is_stiff = 0.0, 1.0, False     # the time span, and whether the system is stiff
    def f_jax(self, t, y, p):              # the right-hand side dy/dt of ONE trajectory
        ...                                # y has shape (dim,), p holds its parameters

problem = Lorenz()
y0, params = ...                           # shapes (n, dim), (n, P): one row per trajectory

# Solve the ensemble; the router picks the engine (engine="tsit5_replay" names one).
result = gradsolve.solve(problem, y0, params)
result.y_final                             # the final states, shape (n, dim)

# Differentiate. final_states(p) is a JAX function of the parameters, so any
# loss built on it can be differentiated with jax.grad.
final_states = gradsolve.grad_closure(problem, y0, params)
loss = lambda p: jnp.sum((final_states(p) - observed) ** 2)   # misfit to the data
gradient = jax.grad(loss)(params)          # shape (n, P): a gradient for every trajectory
\end{verbatim}
\captionof{listing}{The \gradsolve{} interface. A problem is any object exposing
six members: a name, the state dimension, the time span, a stiffness flag, and the
right-hand side \code{f\_jax} of one trajectory. \code{solve} returns the ensemble's
final states; \code{grad\_closure} returns a JAX function of the parameters that any
loss can be differentiated through with \code{jax.grad}. The router selects the
engine, or \code{engine=} names one.}
\label{lst:interface}
\end{minipage}
\end{center}

\section{Benchmarks}\label{sec:benchmarks}

This section reports three main results. Used as a solver alone,
\gradsolve{}'s forward-only kernel is \fwdSamePodDouble{} faster than DiffEqGPU.jl, the reference GPU
ensemble solver. The central one is the reverse-mode gradient: at matched achieved accuracy
\gradsolve{} returns \diffrax{}'s exact gradient at \crossarchRange{} lower cost across three GPU
generations, the margin largest on small non-stiff ensembles and narrowing as they grow, and on stiff
systems a crossing (\stiffRange{}) that favours the replay except at the tightest accuracy. Finally,
the per-gradient saving carries through to complete parameter fits. The subsections that follow give
the evidence, system by system, and the tables collect every measured range.

We evaluate \gradsolve{} on a suite of six systems spanning the workloads it
targets: the Lorenz system in DiffEqGPU.jl's benchmark configuration,
the Van der Pol relaxation oscillator, the stiff Robertson kinetics, the mildly
stiff HIRES system, a flattened dark-matter halo whose stellar orbits are chaotic,
and a Robertson variant whose reaction term is a small neural network;
\ref{app:models} defines all six. Of the baselines we compare against ---
DiffEqGPU.jl, \diffrax{}, torchode and torchdiffeq --- \diffrax{} is the only
one that returns a reverse-mode ensemble gradient, so it is the reference
baseline throughout; torchode and torchdiffeq are compared on the non-stiff
Lorenz gradient. Not every system enters every comparison: the stiff systems appear
where stiffness matters, and the halo and the neural variant carry the fitting
and memory studies. Solvers of different order reach different
accuracy at the same nominal tolerance, so comparing at equal tolerance would
conflate cost with accuracy. Ratios between methods of different order are
therefore read at matched achieved forward-state accuracy: \gradsolve{} runs first and its forward-state error
against a high-accuracy reference is recorded, and the baseline's tolerance is then
lowered by bisection until its own error matches ours, never a higher one
(\ref{app:protocol-matching} states the rule in full). A work--precision sweep plots gradient
cost against achieved error \citep{hairer1993solving}; it is the right comparison
when methods differ in order or error control. Where a ratio is read off such a
sweep, the baseline's cost at our achieved error is interpolated between
its measured points, and no ratio is extrapolated. The forward comparison
(\cref{sec:benchmarks-forward}) instead runs the same Tsit5 pair on both sides at the same nominal tolerance, which makes the achieved accuracies closely comparable rather than identical by construction.

The \diffrax{} baseline runs its default checkpointed adjoint, compiled once and mapped over the ensemble; its step limit and checkpoint count are chosen by measurement for each comparison, and for stiff problems it uses a Newton root finder that measured faster than the library's default. All baselines are timed in reverse mode, and \diffrax{}'s forward mode is timed as well for the systems with one to three inputs per trajectory (\cref{sec:benchmarks-forward-mode}). \ref{app:protocol-baseline} gives every setting. Each per-gradient wall time is the best of three timed calls (five for the higher-order engines) after a warm-up pass that absorbs compilation and, for the
replay engines, the one-time record; five independent repetitions of one such
measurement on the same card, each with a fresh record, vary by \timingSpread{}. The tables and figures time torchode under \code{torch.compile}, its intended fast configuration, which
fuses the per-operation dispatch of \cref{sec:diff-gpu}; run eagerly, out of the box, it is \torchodeRange{}
behind rather than \torchodeCompiledRange{}. torchdiffeq's continuous-adjoint path is timed eagerly
(\torchdiffeqRange{}).

The reverse-mode benchmarks differentiate the loss $\ell = \lVert y(t_1)\rVert^2$
summed over the ensemble, with respect to each trajectory's parameters --- or its
initial state for HIRES, which carries no free rate constant; the fitting studies of
\cref{sec:benchmarks-fitting} instead minimize a data-misfit loss (\ref{app:protocol-fitting}).
Every engine's gradient is checked against a central finite-difference
estimate before its timings are reported. On the stiff Robertson replay the
two agree to about \fdNoiseFloor{}, the resolution of the finite-difference
estimate itself; the non-stiff replay passes the same check on each
system timed below, and the in-kernel adjoint up to $d \approx 12$. These checks
run on the GPU, and \ref{app:gradchecks} repeats the comparison on a CPU.

Measurements are double precision on a single NVIDIA A100, H100, or RTX~4090,
named where the A100's two packagings (PCIe and SXM4) differ; no two frameworks
share a device within one measurement, and \ref{app:versions} lists the machines
and software versions.

Two accounting conventions appear below and are never mixed. Most of the section
reports the \emph{per-gradient cost}: the cost of a single gradient evaluation once
a record exists, the cost that matters when many gradients are taken against the
same mesh. The fitting case studies at the end instead report the \emph{time in the
optimizer loop} of a complete fit once compiled and recorded: compilation and the
one-time record are excluded for both engines, and any mid-fit re-records are
timed. The two answer different questions and are not comparable.

\subsection{Forward-only throughput}\label{sec:benchmarks-forward}

\begin{figure}[!tbp]
  \centering
  \includegraphics[width=\textwidth]{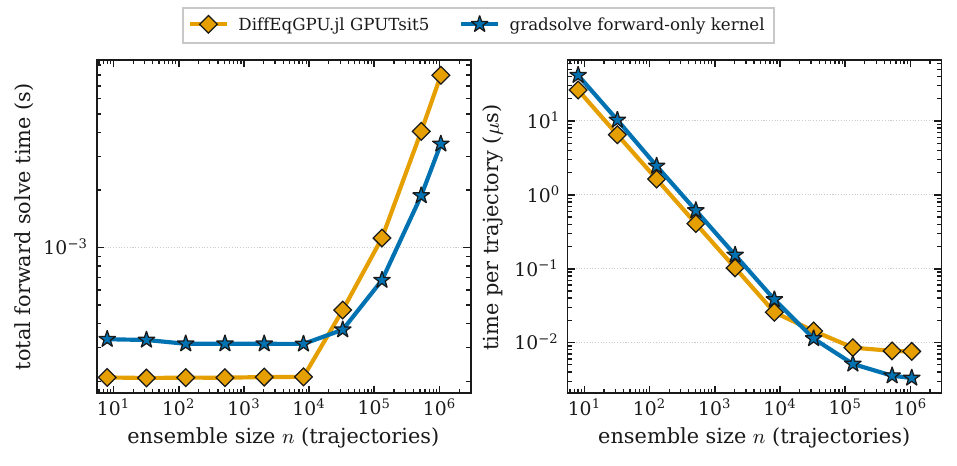}
  \caption{Forward solve time, total (left) and per trajectory (right),
  against ensemble size on the Lorenz system (double precision, matched
  tolerance $\tau_{\mathrm{rel}}=\num{1e-6}$, $\tau_{\mathrm{abs}}=\num{1e-9}$).
  Per-trajectory cost falls with ensemble size as launch overhead
  spreads over more trajectories. The plotted DiffEqGPU.jl sweep ran on a
  different A100 and is shown only for the shape of its scaling; the forward-only
  kernel was swept on an A100-SXM4 at the same tolerances, and the like-for-like
  same-card comparison, in which the forward-only kernel is \fwdSamePodDouble{}
  faster, is given in the text.}
  \label{fig:forward}
\end{figure}

Not every workload requires a gradient, and \cref{sec:method}'s router
directs those that do not, for the registered fields, to the forward-only kernel: one CUDA thread per trajectory, integrating the
Tsit5 pair to a final state. Holding each trajectory's state in registers (\cref{sec:diff-gpu}) means the
whole integration runs as one GPU program, so the fixed cost of starting a GPU
computation is paid once per ensemble rather than once per step. Its
state-dimension limit, shared by every fused kernel, DiffEqGPU.jl's and ours
alike, is measured in \cref{sec:benchmarks-dim}. The comparison against
DiffEqGPU.jl's fused GPUTsit5 kernel \citep{utkarsh2024diffeqgpu} was run with both codes on the same A100-SXM4 card, one after the other, at
matched tolerance ($\tau_{\mathrm{rel}}=\num{1e-6}$, $\tau_{\mathrm{abs}}=\num{1e-9}$)
and $n = \num{1.05e6}$ trajectories:
\gradsolve{}'s kernel is \fwdSamePodDouble{} faster in double precision and
\fwdSamePodSingle{} faster in single precision, the mode DiffEqGPU.jl's own
benchmarks use. \Cref{fig:forward} plots total and per-trajectory solve time against
ensemble size for the two fused kernels; the forward-only kernel is the
fastest integrator we measured on this problem. Against \diffrax{}'s
vectorized adaptive solve, compiled once and timed under the same warm-up
rule at $n = \num{131072}$, the forward-only kernel runs \fwdDiffraxPoint{}
faster per trajectory. Ensemble size is not a practical limit for it: on the same A100-SXM4 card it handles every ensemble size on the grid, reaching \scaleAxisTop{}. A further A100 run sustained \fwdLaneThroughput{} at $n = \num{1.05e6}$, and a capacity run solved \capacityDemo{} in six launches over chunks of the ensemble.

Because a GPU runs the trajectories of an ensemble in groups that advance
together, an ensemble in which a few trajectories need many more steps than the
rest can be held back to the pace of its slowest members. We stress-tested this
by spreading the per-trajectory difficulty of the Lorenz ensemble, so that some
trajectories take far more steps than others. Both fused kernels, ours and
DiffEqGPU.jl's, hold their per-trajectory speed under the spread rather than
slowing to their hardest trajectory, and ours stays \divFwdRange{} faster
throughout (\ref{app:protocol-baseline} gives the settings).

\subsection{Forward-only stiff kernels}\label{sec:benchmarks-stiff-forward}

For stiff problems the library's default GPU kernel is written in a high-level
language, the Warp framework of \cref{sec:method}, and turned into a GPU program
automatically. On the forward-only solve this kernel is \stiffFwdWarpRange{}
slower than DiffEqGPU.jl's stiff kernels, which are written and tuned directly
for the GPU by hand. To find out whether that gap is a limit of the method or
the price of the automatic step, we wrote one stiff kernel directly for the GPU
by hand, for the same second-order method, and compared the three at matched
achieved accuracy on an A100.

The answer depends on the size of the system. On the eight-variable HIRES
system, where each step does a substantial amount of arithmetic, the
hand-written kernel is \cudaRosenHiresVsDiffeqgpu{} faster than DiffEqGPU.jl and
\cudaRosenHiresVsWarp{} faster than the automatic kernel, at every accuracy
measured; the eight-variable state nearly fills the small fast memory each GPU
thread has, so fewer trajectories run at once, and the hand-written kernel is
still faster. On the three-variable Robertson system it matches DiffEqGPU.jl only at the
most demanding accuracy and is up to \cudaRosenRobertsonSlow{} slower at loose
accuracy, because there each solve is very short and a fixed per-trajectory
start-up cost, paid once however few steps follow, dominates. So the slowdown of
the automatically generated kernel is mostly the price of generality, and is
recovered and exceeded by a hand-written kernel where the per-step work is large.
This is a solver-only comparison; \gradsolve{}'s advantage is the reverse-mode
gradient of the next section, which DiffEqGPU.jl does not provide.

\subsection{Reverse-mode gradient cost}\label{sec:benchmarks-reverse}

\begin{figure}[!tbp]
  \centering
  \includegraphics[width=\textwidth]{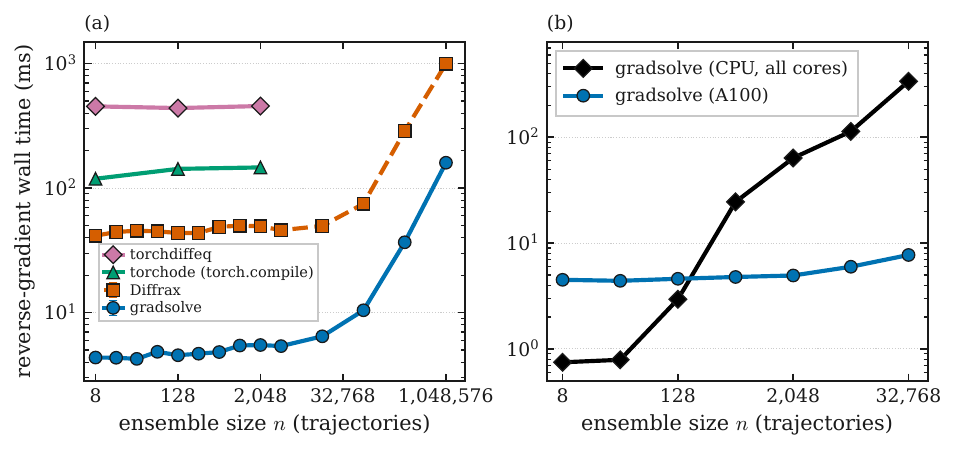}
  \caption{Reverse-mode gradient cost against ensemble size on the Lorenz
  system at matched achieved accuracy (double precision). (a)~On the A100,
  the \recordreplay{} adjoint computes the gradient \revDenseRange{} faster than \diffrax{}'s checkpointed adjoint across the sweep of ensemble
sizes from $n = 8$ to $2048$ and \reverseLargeNRange{} faster out to $n = \reverseSweepTop{}$, and one to
  two orders of magnitude faster than torchode (under \code{torch.compile}) and torchdiffeq, shown at
  representative sizes and timed on a second A100 of the same class. The \diffrax{} and \recordreplay{} points carry an up-whisker to the slowest of their three timed calls, mostly narrower than the marker, the run-to-run spread being \timingSpread{} at worst (\ref{app:protocol-timing}). (b)~The same \recordreplay{} gradient on the CPU of the machine hosting the
A100 and on the A100 itself: the CPU is faster for small
  ensembles, the curves cross between $n = 128$ and $n = 512$, and the GPU
  leads from there upward.}
  \label{fig:reverse}
\end{figure}

\Cref{fig:reverse} plots the cost of one gradient against ensemble size on the
Lorenz system at matched achieved accuracy, for \gradsolve{} and three baselines.
Across the sweep of ensemble sizes from $n = 8$ to $2048$, the \recordreplay{}
adjoint computes the gradient at \revDenseRange{} lower cost than \diffrax{}. At
the three sizes where the PyTorch solvers were also run the margin against
\diffrax{} is \fourSolverDiffrax{}; torchode and torchdiffeq, timed on a second
A100 of the same class (\ref{app:protocol-baseline}), sit \torchodeCompiledRange{} (torchode
under \code{torch.compile}) and \torchdiffeqRange{} behind the \gradsolve{} times of the sweep
(\cref{tab:reverse}). \diffrax{} is the principal baseline; the PyTorch solvers
sit one to two orders of magnitude further back for the dispatch overhead examined
in \cref{sec:whyfast}. The benchmark configuration spreads $\rho$ below the onset
of chaos (\ref{app:models}); a single measurement at the chaotic value $\rho = 28$
gives \lorenzChaoticPoint{} at $n = 2048$.

The three baselines differ in what they differentiate. \diffrax{} and torchode,
like \gradsolve{}, form a \emph{discrete} adjoint (\cref{sec:diff-admodes}), but
of the adaptive solve itself rather than of a recorded replay. \diffrax{} does so
through its checkpointed adjoint, which stores checkpoints of the forward loop and
recomputes the segments between them when there are fewer checkpoints than steps,
the online checkpointing of \citet{stumm2010new} as implemented in Equinox \citep{kidger2021equinox}. In most of the measurements reported here the fastest of the three settings placed a checkpoint on every attempted step, so nothing was recomputed and the cost is that of the differentiated loop itself. torchode
\citep{lienen2022torchode} does so with a separate step-size controller for each
trajectory, while still advancing the whole batch in one tensor operation.
torchdiffeq \citep{chen2018torchdiffeq} instead forms a \emph{continuous} adjoint: it
derives the adjoint differential equation of the original system and solves it
backward in time with the Dormand--Prince pair \citep{dormand1980family}, so its
gradient is that of the exact solution, discretized afresh, rather than of the
steps the forward solve took.

DiffEqGPU.jl is compared on the forward side only. Inside its fused kernels it
differentiates in forward mode only, and for reverse mode its documentation routes
ensembles through its array-based path (\code{EnsembleGPUArray}), the lockstep
design of \cref{sec:diff-gpu}. On the versions listed in \cref{tab:versions}
(Julia~1.12.6, DiffEqGPU.jl~3.15.3, CUDA.jl~6.2.1), neither
path returned a reverse-mode ensemble gradient. We tried three routes, each of which failed to compile or launch: the array path under \code{SciMLSensitivity} with Zygote, the reverse route its documentation recommends, which raised a \code{DimensionMismatch} projecting the gradient of the per-trajectory parameters; and the fused kernel under Enzyme and under Zygote, which raised an \code{InvalidIRError} and a \code{MethodError} during GPU compilation. The array path supports reverse-mode adjoints in general, so this reflects the per-trajectory-parameter ensemble on the tested versions rather than a limitation of the approach in principle.

The GPU is not the faster device at every ensemble size, so we measured where
it overtakes the CPU. Panel~(b) of \cref{fig:reverse} runs the same gradient on the CPU of the machine hosting the A100 (all cores)
and on the A100. The CPU is faster for the smallest ensembles, where the GPU's
fixed overheads dominate its runtime; the curves cross between $n = 128$ and
$n = 512$, and from $n = 512$ upward the GPU leads, reaching \cpuGpuCrossoverTop{} at
$n = \num{32768}$.

Panel~(a)'s \diffrax{} and \gradsolve{} curves are nearly flat over the same range of ensemble sizes:
while the ensemble is too small to occupy the device, fixed per-step costs
dominate both runtimes, and the ratio between them is a ratio of those fixed
costs. The sweep therefore continues to $n = \reverseSweepTop{}$; from
$n = 4096$ upward the advantage reads \reverseLargeNRange{}; the replay's own
time only begins to grow with $n$ above about $6\times10^{4}$ trajectories,
so a user with ensembles above $10^{5}$ trajectories should expect the lower end of that range.

\begin{figure}[!tbp]
  \centering
  \includegraphics[width=\textwidth]{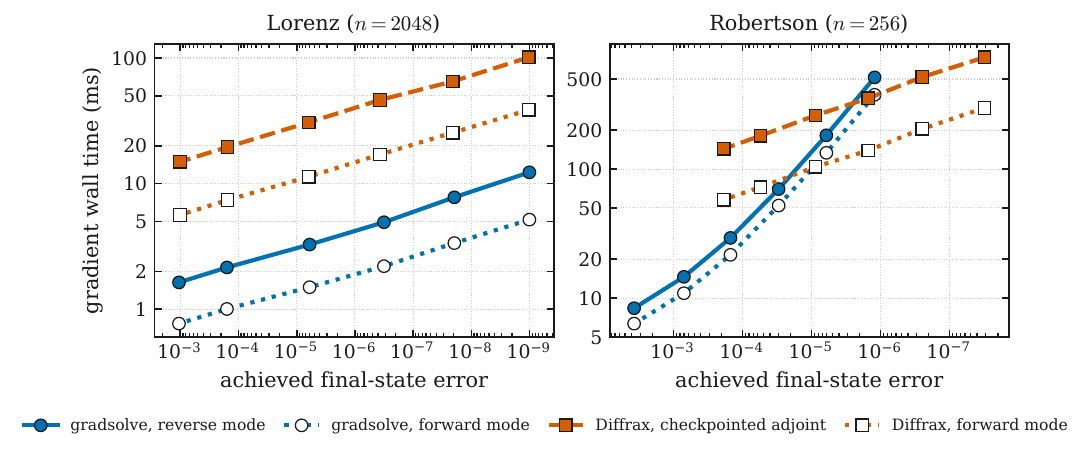}
  \caption{Gradient work--precision on the Lorenz ($n = 2048$, left) and Robertson ($n = 256$, right) systems (A100, double precision): time for one gradient against achieved final-state error, so the engines are compared at matched achieved accuracy. Filled markers are reverse mode (the \recordreplay{} adjoint and \diffrax{}'s checkpointed adjoint), open markers forward mode (the same replay and \diffrax{}'s \code{ForwardMode}, one tangent pass per input). On Lorenz the four curves are parallel and the replay sits below \diffrax{} in both modes. On Robertson the replay's curves are steeper, because its second-order method needs more steps than the fifth-order baseline as accuracy tightens: they cross \diffrax{}'s checkpointed adjoint at an achieved error between \num{1e-6} and \num{1e-5} and its forward mode near \num{1e-5}. All four arms were timed in one session, whose two reverse curves reproduce the ranges quoted in the text within the run-to-run spread.}
  \label{fig:workprec}
\end{figure}

A second view holds the ensemble size fixed and sweeps the tolerance. \gradsolve{} and \diffrax{} advance the same Tsit5 pair, so both sweep the same tolerance range; \cref{fig:workprec} plots gradient time against achieved final-state error from $\tau_{\mathrm{rel}}=\num{1e-3}$ to $\num{1e-8}$, and the same figure also carries the forward-mode arms of \cref{sec:benchmarks-forward-mode}. Read at matched achieved accuracy within the curves' overlap by
the interpolation rule above, the replay computes the gradient at
\revwprange{} lower cost across the range, the advantage declining mildly toward the tightest tolerance.

\begin{figure}[!tbp]
  \centering
  \includegraphics[width=\textwidth]{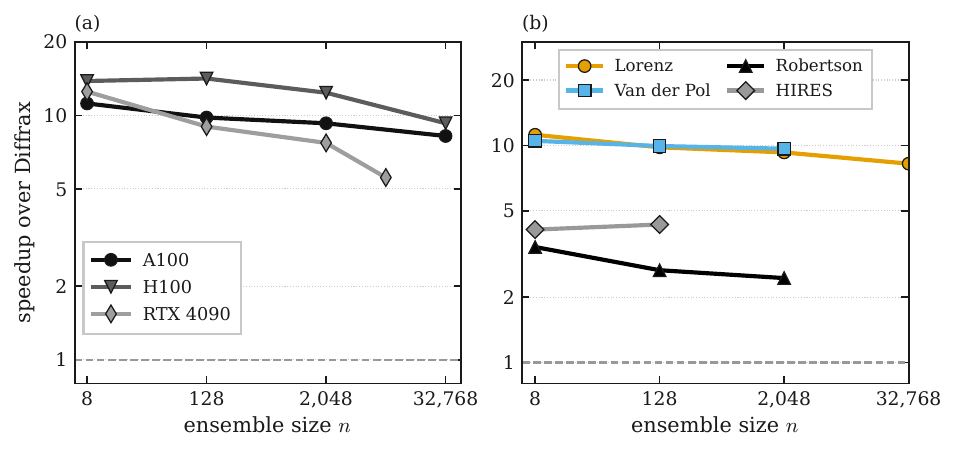}
  \caption{Matched-accuracy gradient speedup over \diffrax{} against
  ensemble size (double precision). (a)~Lorenz on three GPU generations:
  the advantage spans \crossarchRange{} on the three cards and declines with
  ensemble size on each, remaining above the parity line (dashed). (b)~Four
  systems of the suite on the A100: the non-stiff Lorenz and Van der Pol systems and the stiff Robertson and HIRES systems, all above parity at the
  accuracy swept here. HIRES differentiates with respect to its initial
  state, since it has no free parameter.}
  \label{fig:crossarch}
\end{figure}

A speedup measured on one device can reflect that device as much as the
method, so we repeated the Lorenz comparison on three GPU generations.
\Cref{fig:crossarch} plots the speedup over \diffrax{} against ensemble size on each card. The advantage stays within
\crossarchRange{} on the A100, the H100, and the RTX 4090, is highest on the
H100, and falls from its small-ensemble value on all three, most steeply on
the RTX 4090, where it spans \crossarchFourNinety{} from $n = 8$ to $n = 8192$. The A100 and H100 sweeps
here reach $n = \num{32768}$; across its own sweep the RTX 4090's gradient
times are comparable to the A100's at the same ensemble sizes, so the card's much lower double-precision throughput does
not set these ratios.

The advantage is not specific to the Lorenz system. Panel~(b) of \cref{fig:crossarch}
repeats the matched-accuracy comparison across the suite, one speedup
curve per system: the Van der Pol oscillator returns \revVdPRange{}, close
to the Lorenz range, and the two stiff systems, taken up next, stay above
parity throughout this sweep, Robertson at \stiffSweepRange{} and HIRES at
\hiresLowOrder{}. 

The registered fields can also run in single precision. That changes the
gradient cost only where the kernel dominates the wall time: at
$n = \num{131072}$ the Lorenz gradient in single precision runs in
\singleLorenzGradRatio{}\x{} its double-precision time, and
about nine tenths of that at smaller $n$. The stiff Robertson gradient shows
essentially no change, taking \singleRobertsonGradRange{}\x{} its double-precision
time at every ensemble size measured, from 256 to \num{16384} trajectories, because there
the kernel does not dominate the wall time. The single-precision gradients agree with the
double-precision ones to \singleGradAgreement{} relative.
Taken together, the views agree: at matched accuracy the replay returns the same
gradient \diffrax{} does --- run on a shared mesh the two agree to \sharedMeshLorenz{}
(\ref{app:gradchecks}) and computes it several times faster, by a margin that is largest on small
non-stiff ensembles and shrinks smoothly as the ensemble grows, without inverting anywhere non-stiff.

\subsection{Forward mode for few inputs}\label{sec:benchmarks-forward-mode}

These gradients can also be obtained in forward mode. On the Lorenz, Van der
Pol and Robertson ensembles each trajectory is differentiated with respect to one
to three of its own parameters. No trajectory's solution depends on another
trajectory's inputs, so a single forward-mode pass can change the same input in
every trajectory at once and still return each trajectory's own derivative; the
number of passes is therefore set by the inputs of one trajectory, not by the size
of the ensemble (\ref{app:protocol-baseline}). HIRES is not timed this way, because
its gradient is taken with respect to an eight-component initial state, and neither
is the neural right-hand side, whose thousands of parameters leave reverse mode as
the only option. \diffrax{} provides forward mode as its \code{ForwardMode} option,
a plain loop with no checkpoints, and its documentation recommends it when inputs
are few; the replay, being an ordinary JAX program, can be differentiated in
forward mode without any change. We therefore timed both engines in forward mode
as well, on the same batches and at the same matched tolerances, in a separate
session on the A100. \Cref{fig:workprec} shows all four arms, and
\cref{tab:forward} collects the ranges, including the ensemble-size sweeps and the
by-name engines of \cref{sec:benchmarks-highorder} against Dopri8 and Kvaerno5 in
forward mode.

On Lorenz, \diffrax{}'s forward mode runs \dfxFwdOverRevLorenz{} faster than its
checkpointed adjoint, so it is the setting a \diffrax{} user should choose for this
problem. Against it, the reverse-mode replay is still \fwdModeLorenzRange{} faster
at matched achieved accuracy across the tolerance range, and the replay in forward
mode \fwdReplayLorenzRange{} faster; the Van der Pol oscillator and the
ensemble-size sweeps behave the same way (\cref{tab:forward}). Mode for mode the
advantage is close: \revwprange{} in reverse mode against the checkpointed adjoint,
\fwdReplayLorenzRange{} in forward mode against forward mode. The stiff Robertson
system is the exception. Its forward mode also runs \dfxFwdOverRevRobertson{} faster
than its adjoint, and against it the second-order Rosenbrock23 replay holds
\fwdModeRobertsonRange{} across the accuracy range (\fwdReplayRobertsonRange{} in
forward mode): above parity at loose accuracy, and below it from an achieved error
of about \num{1e-5} onwards, where the fifth-order baseline needs far fewer steps.
The fifth-order Rodas5P replay of \cref{sec:benchmarks-highorder} recovers
\fwdModeRodasRobertson{} at its measured point. The replay is faster in forward mode
as well, where neither engine stores intermediate values, so its saving comes
from the loop structure itself (\cref{sec:whyfast}).

\begin{table}[!htb]
  \centering\footnotesize
  \caption{Matched-accuracy speedup over \diffrax{}'s forward mode
  (\code{ForwardMode}, one tangent pass per input) for the systems with one to three
  inputs per trajectory, with the replay differentiated in reverse mode and in forward
  mode. All arms were timed in one A100 session. A single value is quoted where a range
  collapses at one decimal or, in the by-name rows, where one ensemble size ($n = 1024$)
  was measured.}
  \label{tab:forward}
  \setlength{\tabcolsep}{4pt}
  \begin{tabular}{llcc}
    \toprule
    Comparison & Baseline & Replay in reverse mode & Replay in forward mode \\
    \midrule
    Lorenz, tolerance range & \diffrax{} Tsit5 & \fwdModeLorenzRange{} & \fwdReplayLorenzRange{} \\
    Lorenz, $n = 8$--\num{262144} & \diffrax{} Tsit5 & \fwdModeLorenzSweep{} & \fwdReplayLorenzSweep{} \\
    Van der Pol, size sweep & \diffrax{} Tsit5 & \fwdModeVdPRange{} & \fwdReplayVdPRange{} \\
    Robertson, loose to tight accuracy & \diffrax{} Kvaerno5 & \fwdModeRobertsonRange{} & \fwdReplayRobertsonRange{} \\
    Robertson, size sweep, loose accuracy & \diffrax{} Kvaerno5 & \fwdModeRobertsonSweep{} & \fwdReplayRobertsonSweep{} \\
    \midrule
    Lorenz, Verner pair (order 7) & \diffrax{} Dopri8 & \fwdModeVernLorenz{} & \fwdReplayVernLorenz{} \\
    Robertson, Rodas5P engine & \diffrax{} Kvaerno5 & \fwdModeRodasRobertson{} & \fwdReplayRodasRobertson{} \\
    \bottomrule
  \end{tabular}
\end{table}

\subsection{Stiff systems}\label{sec:benchmarks-stiff}

Adaptive step control does the most work on stiff systems, which therefore
test whether removing it from the differentiated computation remains an
advantage. On the
Robertson kinetics \citep{robertson1966solution, hairer1996solving}, the stiff replay, running on the recorded Rosenbrock23 mesh, is compared
against \diffrax{}'s higher-order Kvaerno5 baseline, a singly diagonally implicit Runge--Kutta method with an explicit first
stage, designed for stiff problems \citep{kvaerno2004singly}. How much the replay saves
depends on the accuracy demanded. Over the range of achieved error the two sweeps share (\stiffOverlap{}
decades), the advantage reads \stiffRange{}. It is
\stiffRangeMax{} at loose accuracy and reaches parity at an achieved error
between \num{1e-6} and \num{1e-5}; at the tightest point it falls slightly
below parity, because there the fifth-order baseline needs far fewer steps
than the second-order replay to reach a given error. For tight accuracy the stiff engine is instead the fifth-order Rodas5P replay (\cref{sec:benchmarks-highorder}). Against \diffrax{}'s forward mode, which suits this three-parameter system, the margins are smaller (\cref{sec:benchmarks-forward-mode}). A companion sweep in ensemble size at one tolerance, at the loose end of that
range, gives \stiffSweepRange{} on the A100, declining from $n = 8$ to
$n = 2048$ (the Robertson curve of \cref{fig:crossarch}(b)).

HIRES
\citep{schafer1975hires, hairer1996solving}, an eight-dimensional stiff
system, also sits above parity. Its ensemble carries no free rate constant
(\ref{app:models}), so the gradient is taken with respect to the initial
state; at matched achieved accuracy the stiff replay computes it \hiresLowOrder{}
faster than \diffrax{} across $n = 8$ to $128$. The higher-order Rodas5P
engine widens the HIRES margin further (\cref{sec:benchmarks-highorder}).
On stiff systems, then, the replay helps up to moderate accuracy, and at the tightest accuracy
\diffrax{} is the better choice; the fifth-order Rodas5P replay moves that boundary out where it is
needed.

\subsection{Dimension limit of the fused kernels}\label{sec:benchmarks-dim}

\begin{figure}[!tbp]
  \centering
  \includegraphics[width=0.83\textwidth]{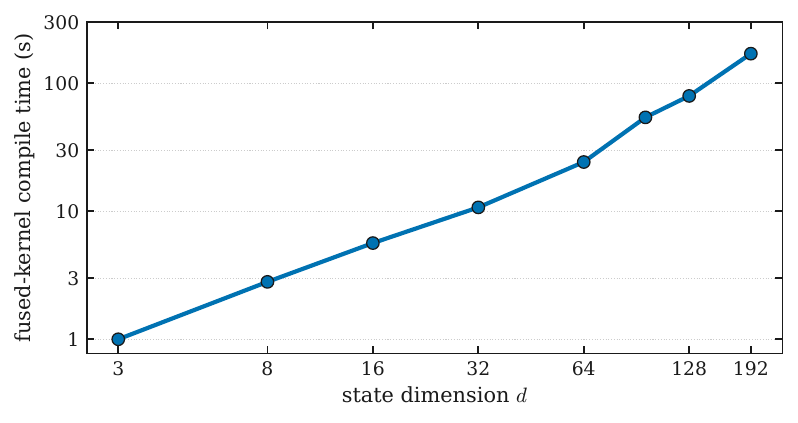}
  \caption{Fused-kernel compile time against state dimension on the linear
  family of \ref{app:models} ($n = 8192$, A100). The next dimension probed, $d = 316$,
  compiles in \SI{322}{\second} but fails at launch and is not plotted. The family's right-hand side is
  trivial, so the axis isolates the compilation limit, not speed.}
  \label{fig:dimsweep}
\end{figure}

The fused kernels have a state-dimension limit, and it appears abruptly, at launch, rather than as a gradual loss of speed. We measured it on the linear family of \ref{app:models}, whose state
dimension is free (\cref{fig:dimsweep}). For the fused recorder, compile time grows from about \SI{1}{\second} at $d = 3$ to \SI{322}{\second} at $d = 316$, where the kernel compiles but fails at launch. The router's
fallback to the general path at $d = 64$ is a conservative empirical policy: it sits far inside
this launch boundary, leaving margin. The failure at $d = 316$ is a launch-time error --- the GPU
refusing the kernel for want of per-thread resources, the signature of register exhaustion
(\cref{sec:diff-gpu}) --- and the figure plots compile time rather than a register count, so it
locates the boundary without profiling its cause.

\subsection{Higher-order engines}\label{sec:benchmarks-highorder}

The two higher-order methods of \cref{sec:method}, the seventh-order Verner
pair and the fifth-order Rosenbrock method Rodas5P, were timed under the same matched-accuracy procedure in a separate run on an
A100-SXM4 at $n = 1024$
(\cref{tab:reverse}). Neither engine is what the router picks for a
registered field: both were measured against \diffrax{} rather than against
the fused engines the router prefers, so the ratios below are what a user
obtains by asking for the method by name. Rodas5P is also the general stiff engine the router selects for gradients
outside the registered fields. The Verner engine computes the
Lorenz gradient \mevVernLorenz{} faster than \diffrax{}'s Dopri8, an eighth-order explicit method and the
faster of the two \diffrax{} explicit methods timed on this problem; the
Rodas5P engine is \mevRodasRobertson{} faster on Robertson and
\mevRodasHires{} faster on HIRES than Kvaerno5. The first two differentiate
with respect to the ensemble's rate parameters, HIRES with respect to the
initial state, as in \cref{sec:benchmarks-stiff}. On this system the
higher-order Rodas5P replay attains a larger margin than the Rosenbrock23
replay, at the tighter accuracy it reaches (\num{1.3e-7}).

The one-time record is excluded here as everywhere in this section and reported separately. At this ensemble size the general recorder records the three meshes on the A100 in \recordDeviceLorenz{} (Lorenz), \recordDeviceRobertson{} (Robertson) and \recordDeviceHires{} (HIRES). Against the per-gradient saving over \diffrax{} on the same three comparisons, the device record costs \recordPaybackRange{} of one gradient's saving, so it is repaid before the first gradient is complete: counting the record, \gradsolve{}'s first gradient on each of the three still costs less than a single \diffrax{} gradient, which carries no such record phase. A fourth system, the linear family at $d = 32$, differentiates with respect to
its initial state, and its loss is a pure rescaling of that state, so the
central-difference signal falls to the resolution of the finite-difference
estimate itself and cannot discriminate the gradient. A linear system, however,
has a closed-form sensitivity; the replay's gradient agrees with it to
\num{1.6e-7} relative, the order of the method's own forward error. Validated
that way, the Verner engine computes this gradient \mevVernLinear{} faster than
\diffrax{}'s Dopri8, the widest of the four margins, because the linear field's
step is so cheap that the differentiated loop dominates the cost.

The advantage therefore does not depend on the integrator: the same simpler backward sweep
carries the seventh-order Verner pair and the fifth-order Rodas5P as it does the second-order Tsit5 and
Rosenbrock23.

A further control, which holds \diffrax{} to a fixed mesh of our own step count, is
reported with the analysis of \cref{sec:whyfast}.

\subsection{Fitting case studies}\label{sec:benchmarks-fitting}

To see whether the per-gradient saving carries through to a complete
calibration, we also timed \gradsolve{} inside fitting loops, where Adam \citep{kingma2015adam}
consumes the gradient step after step until a physical parameter is
recovered. These are times in the optimizer loop, which exclude compilation and the one-time record for both engines; we run $B$ concurrent fits, $B \in \{1, 8, 64, 256\}$, each recovering its own parameter, with both engines under the identical optimizer (\ref{app:protocol-fitting} gives the protocol). The recorded mesh is treated as data for the duration of a fit. A companion measurement re-solves adaptively at parameter snapshots along the
full fits of all three systems on a CPU, at their full \num{400} updates and at
the batch sizes the fits reach ($B = 256$ for the Lorenz and galactic fits,
$B = 64$ for the stiff Robertson fit, the largest at which it converges).
Relative to the replay of the starting mesh, the fresh solve's final state moves
by at most \num{2.1e-6} at every snapshot, even where the Robertson rates
traverse a decade. The two gradients agree in direction to within $0.05^{\circ}$
while the optimizer is descending --- while the fresh-solve gradient still keeps
at least a hundredth of its initial norm; once a fit converges the gradient falls
by four to nine orders of magnitude on both meshes and the direction of that
vanishing gradient is no longer defined, but the optimizer has by then stopped
moving and every fit has already recovered its parameter on the starting mesh.
The starting mesh therefore remains adequate throughout a fit. 

\begin{figure}[!tbp]
  \centering
  \includegraphics[width=\textwidth]{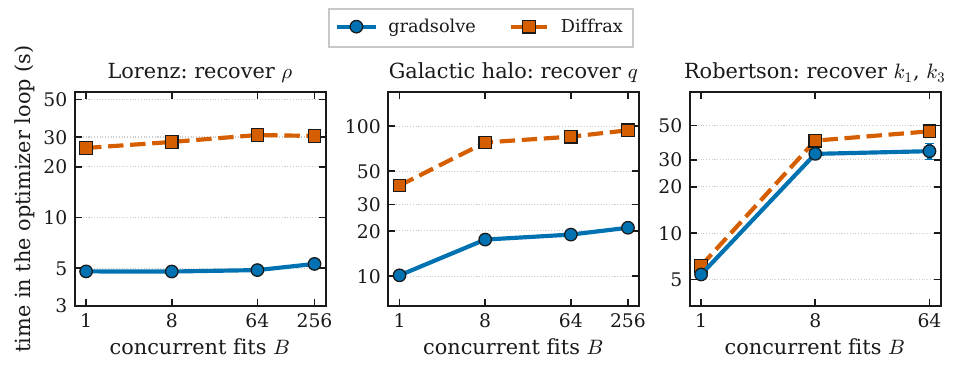}
  \caption{Time in the optimizer loop against the number of concurrent fits $B$ at matched
  accuracy (A100), recovering the Lorenz Rayleigh parameter $\rho$ (left), the
  galactic halo flattening $q$ (center), and the Robertson rate constants
  (right). The lower gradient cost translates into a faster fit in every
  panel. Error bars are the standard deviation over the timed repetitions; the
  run-to-run spread is under about $3\%$ at almost every point (up to about
  $12\%$ at the noisiest), so the bars mostly fall within the markers. The
  Robertson $B = 256$ point is absent because neither engine reached the loss
  tolerance there.}
  \label{fig:fitting}
\end{figure}

The simplest study fits the flattening $q$ of a galactic dark-matter halo,
modeled by the logarithmic potential
$\Phi = \tfrac{1}{2} v_c^2 \ln(R_c^2 + x^2 + y^2 + z^2/q^2)$
\citep{binney2008galactic}. The flattening is not observed directly; it is
inferred from how stars move through the potential, and for $q < 1$ many of those
orbits are chaotic, a setting in which a gradient through the solve is particularly valuable. Every fit was recovered by both engines at every value of $B$, so the
comparison is a pure timing contrast: the time in the optimizer loop favors
\gradsolve{} by \fitGalactic{}
across the four values of $B$
(\cref{fig:fitting}). In absolute terms, recovering $q$ for 256 concurrent fits over 400 updates took \fitGalacticSeconds{}. Recovering the Lorenz Rayleigh parameter $\rho$
from noisy final-state observations gives the largest margin,
\fitLorenz{}, narrowest at $B = 1$ and widest at $B = 64$, easing
slightly at $B = 256$.
Recovering the Robertson rate constants $k_1$ and $k_3$ through the stiff replay holds a
smaller, flatter advantage, \fitRobertson{}, for $B \le 64$; at
$B = 256$ neither engine reached the loss tolerance within its step cap,
so that combination reports no ratio. All three studies land below the per-gradient ranges because the
optimizer bookkeeping, common to both engines, is paid on
every update and dilutes the per-gradient saving. The Lorenz and galactic fits use the general path of \cref{sec:method}, the path taken by any right-hand side without a registered fused kernel (the one-time record is outside the timed loop), so their ratios are the ones a user's own model obtains on this hardware; the Robertson fit uses the fused
stiff recorder.

Finally, we placed a small neural network (a multilayer perceptron,
\ref{app:models}) inside the Robertson system's autocatalytic term and
trained it. Training uses
the record-and-replay gradient, which agreed with a central finite-difference
estimate to \udeFdReplay{}; the loss fell from \num{8.9e7} to \num{7.3e5} over
20 optimizer updates, so the mechanism carries over when the right-hand side is itself a
trainable function. This is a capability demonstration --- the record-and-replay gradient trains
a neural right-hand side, and, with the in-kernel adjoint below, does so from a compact record ---
not a claim of learning or generalization, which a held-out evaluation would be needed to establish.

\begin{figure}[!tbp]
  \centering
  \includegraphics[width=\textwidth]{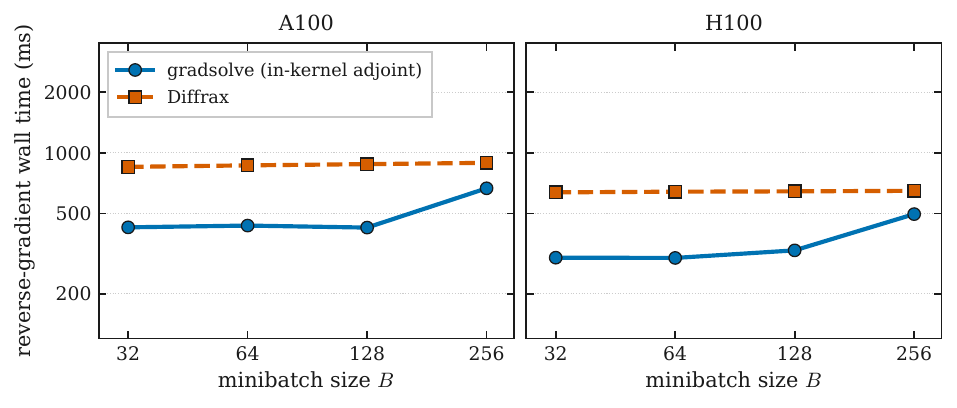}
  \caption{Per-step gradient time against minibatch size $B$ for the
  Robertson system with a neural reaction term, on the A100 and the H100, with
  \diffrax{}'s tolerance fixed at the value matching the replay's achieved
  error. \gradsolve{}'s in-kernel adjoint is faster at every minibatch size
  measured, by
  a margin that is about $2\x$ at the smaller batch sizes and narrows at $B = 256$.}
  \label{fig:udebscaling}
\end{figure}

The same neural right-hand side also runs in the in-kernel adjoint of \cref{sec:method}, whose backward sweep we verified separately against finite differences to \udeFdFused{}; this is the engine whose compact record gives the memory advantage. For this comparison \diffrax{}'s tolerance is fixed in advance at the value that matches the replay's achieved error, rather than bisected at each minibatch size (the number of trajectories per training step, $B$). The margin is narrower than on the mechanistic systems: the in-kernel adjoint's per-update gradient is \udeSmallBatch{} faster than \diffrax{}'s checkpointed baseline at $B = 32$ and $B = 64$ on the A100, narrowing to \udeLargeBatch{} at $B = 256$ on both the A100 and the H100 (\cref{fig:udebscaling}).

\Cref{tab:reverse} collects the per-gradient ranges against \diffrax{}'s checkpointed adjoint, \cref{tab:forward} those against its forward mode, and \cref{tab:fitting} the fitting case studies.

\begin{table}[tb]
  \centering\footnotesize
  \caption{Matched-accuracy speedup ranges over \diffrax{} and, where
  noted, over the PyTorch solvers, reported as measured minimum--maximum
  ranges. Every row reports the cost of a single gradient against an existing record.
  The last three rows time the two higher-order engines at a single
  ensemble size ($n = 1024$), so each carries one value rather than a range.
  The two PyTorch rows divide times measured on a second A100 of the same class by \gradsolve{} times from the sweep; torchode is timed under \code{torch.compile}, its intended fast configuration (eager, out of the box, it is \torchodeRange{}), torchdiffeq's continuous adjoint eagerly. }
  \label{tab:reverse}
  \begin{tabular}{llc}
    \toprule
    Comparison & Baseline & Speedup \\
    \midrule
    Lorenz, tolerance range (work--precision) & \diffrax{} & \revwprange{} \\
    Lorenz, ensemble-size sweep $n = 8$--$2048$ & \diffrax{} & \revDenseRange{} \\
    Lorenz, $n = 4096$--\reverseSweepTop{} & \diffrax{} & \reverseLargeNRange{} \\
    Lorenz, sizes run with all four solvers & \diffrax{} & \fourSolverDiffrax{} \\
    Van der Pol, ensemble-size sweep & \diffrax{} & \revVdPRange{} \\
    Lorenz, sizes run with all four solvers & torchode (\code{torch.compile}) & \torchodeCompiledRange{} \\
    Lorenz, sizes run with all four solvers & torchdiffeq & \torchdiffeqRange{} \\
    Lorenz, three architectures (A100 / H100 / RTX 4090) & \diffrax{} & \crossarchRange{} \\
    Robertson (stiff), loose to tight accuracy & \diffrax{} & \stiffRange{} \\
    Robertson (stiff), ensemble-size sweep, A100 & \diffrax{} & \stiffSweepRange{} \\
    Robertson (stiff), ensemble-size sweep, RTX 4090 & \diffrax{} & \stiffSecondArch{} \\
    HIRES (stiff) & \diffrax{} & \hiresLowOrder{} \\
        \midrule
    Lorenz, Verner pair (order 7) & \diffrax{} Dopri8 & \mevVernLorenz{} \\
    Robertson, Rodas5P engine & \diffrax{} Kvaerno5 & \mevRodasRobertson{} \\
    HIRES, Rodas5P engine & \diffrax{} Kvaerno5 & \mevRodasHires{} \\
    \bottomrule
  \end{tabular}
\end{table}

\begin{table}[tb]
  \centering\footnotesize
  \caption{Fitting case studies: speedup in the optimizer loop over
  \diffrax{} at matched achieved accuracy (time in the optimizer loop; compilation and the one-time record excluded for both engines). Lorenz and
  galactic fits run 400 updates; the Robertson fit stops
  at a loss tolerance and is valid for $B \le 64$.}
  \label{tab:fitting}
  \begin{tabular}{lc}
    \toprule
    Study & Speedup range \\
    \midrule
    Galactic halo $q$ recovery & \fitGalactic{} \\
    Lorenz $\rho$ recovery & \fitLorenz{} \\
    Robertson rate recovery, $B \le 64$ & \fitRobertson{} \\
    \bottomrule
  \end{tabular}
\end{table}

\section{Where the speedup comes from and when to expect it}\label{sec:whyfast}

The benchmarks of \cref{sec:benchmarks} establish how large the speedup is;
this section identifies where it comes from. Three explanations are plausible: uneven step counts across the ensemble, the fused recorder, and a difference in the differentiated computation itself. Measurements rule out the first two as the main cause; a control experiment then separates the third into the missing step-size controller and the fixed-length loop, and traces the saving to the loop. Knowing this tells a user when to expect the method to help.

\emph{Uneven step counts.} The obvious candidate is the waiting described in \cref{sec:diff-gpu}.
When trajectories are batched in lockstep, every trajectory steps until
the slowest one finishes, so a method that frees each trajectory to take only
its own steps would appear to gain exactly that wasted wait. We tested
this directly. On the Lorenz ensemble used for the reverse-mode comparison
of \cref{sec:benchmarks-reverse}, the slowest trajectory accepts only
\dispersionRange{} times as many steps as the average
(\cref{tab:dispersion}); freeing it from the lockstep can therefore save only a
factor of that order, whereas the measured gradient advantage is \revwprange{},
several times larger. A second solver confirms that the spread is real: reading the same ensemble
through torchode's own step counters reproduces it (\cref{tab:dispersion}).
Uneven step counts therefore contribute a small factor but do not explain the
gap.

\begin{table}[tb]
  \centering
  \caption{Accepted-step counts across the Lorenz timing ensemble
  ($\rho \in [0, 21]$) of \cref{fig:reverse}, recorded at $\tau_{\mathrm{rel}}=\num{1e-6}$,
  $\tau_{\mathrm{abs}}=\num{1e-9}$. The second row reads the same ensemble through torchode's step counters; the linear family of \ref{app:models}
  needs the same number of steps for every trajectory.}
  \label{tab:dispersion}
  \begin{tabular}{lccc}
    \toprule
    Ensemble and step counter & mean steps & max steps & max\,/\,mean \\
    \midrule
    Lorenz, fused recorder          & 36.7 & 49 & 1.34 \\
    Lorenz, torchode's step counters  & 39   & 52 & 1.33 \\
    Linear family ($d=3$)           & \multicolumn{2}{c}{equal for all trajectories} & 1.00 \\
    \bottomrule
  \end{tabular}
\end{table}

\begin{figure}[!tbp]
  \centering
  \includegraphics[width=0.83\textwidth]{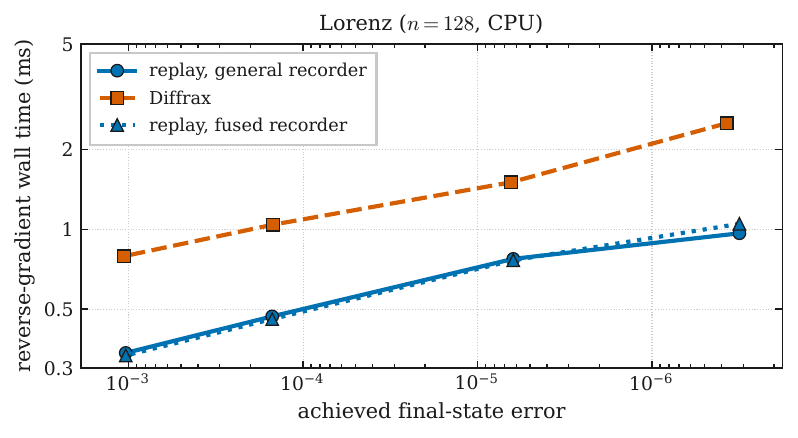}
  \caption{Per-gradient cost against achieved final-state error on the
  Lorenz system ($n = 128$, double precision, CPU). Both \gradsolve{} curves run the same
  differentiated JAX replay and differ only in how the mesh was recorded, by
  the general recorder or by the fused recorder; recording is excluded from the
  per-gradient time. The two therefore track each other closely, and both sit
  below \diffrax{}'s checkpointed adjoint across the tolerance range. The gap
  to \diffrax{} thus measures the effect of differentiating a fixed-step
  replay rather than of any GPU kernel.}
  \label{fig:whyfast}
\end{figure}

\emph{The fused recorder.} In a CPU experiment at $n = 128$, where no GPU effect can contribute, the replay on the general path computes gradients \scanWpRange{} faster than \diffrax{} and the replay with the fused recorder \fusedWpRange{} faster. The comparison is at equal tolerance, which here means equal accuracy because the same Tsit5 pair runs on both sides; the two curves track each other closely (\cref{fig:whyfast}). Both differentiate the same fixed-step replay, so the recorder is not the
source; the GPU ranges of \cref{sec:benchmarks} are those of the same replay. The
same comparison on the CPU gives only \scanWpRange{}, against
\revDenseRange{} on the A100 (\cref{fig:reverse}), so roughly a factor of
four of the GPU advantage is specific to how the GPU executes the two program
forms, which the control below isolates.

\emph{The absence of the controller.} A further control asks how much of each
difference comes merely from not having to choose the steps. For it, \diffrax{} was
run on a fixed mesh with our own step count, hence with no controller and no
rejected trials, on the three by-name comparisons of
\cref{sec:benchmarks-highorder}. Removing the step-size choice alone does not
account for the differences: on Lorenz and Robertson the fixed-mesh run is slower
than the matched adaptive one, and on HIRES it is faster but reaches an error of
\num{5.9e-4} against our \num{1.3e-7}. The HIRES speedup is therefore partly the
fixed mesh rather than the loop, by an amount this run cannot fix. These runs are
not accuracy-matched and use \diffrax{}'s own methods, Dopri8 and Kvaerno5, whose
steps cost more than the Verner and Rodas5P steps of the replay, on a mesh of the
same count but not the recorded shape (\ref{app:protocol-baseline}), so their
ratios are rough comparisons rather than matched measurements. Measured against the replay itself, the fixed-mesh runs cost
\frozenMeshRatios{} as much as the replay's gradient on Lorenz, Robertson, and HIRES
respectively. That remaining gap mixes the loop form with the per-step cost of the
different methods. A last control removes the per-step cost too: the same Tsit5 method on the
recorded mesh, stepped through \diffrax{}'s \code{StepTo} controller so that only the loop form
differs. On a homogeneous Lorenz ensemble --- one shared recorded mesh, so no step-count spread ---
the replay's fixed-step scan computes the gradient \steptoLoopFormGpu{} faster than the \code{StepTo}
loop on the A100 at the same accuracy (their gradients agree to \num{2e-16}), which is essentially the
whole advantage of \cref{sec:benchmarks-reverse}. On the CPU the scan holds no such advantage over the
\code{StepTo} loop --- it is in fact slightly slower on the larger ensembles --- so the advantage is the
GPU's execution of a fixed-length scan rather than a bounded, checkpointed while-loop, the same factor of
four the fused-recorder comparison left open.

The fixed mesh leaves the differentiated \emph{loop} in place, and its form is
the part these controls point to: a fixed-length chain compiles to one regular program
whose backward sweep is one linear pass, whereas an adaptive loop differentiated in
place is bounded, checkpointed, and advanced through every trial of every
trajectory. The replay decides once, when it is compiled, whether to keep each intermediate value or recompute it, and then does the same at every step. The adaptive loop, even with a checkpoint at every step so that nothing is recomputed, must still write its intermediate values to memory and read them back from positions it only knows at run time.

\emph{Where the time goes.} In the Lorenz gradient computation with the fused
recorder on the A100 ($n = \num{262144}$, double precision), recording is inexpensive: of the cost of recording once and taking
one gradient, the record is $11$--$15\%$ and the backward sweep about $80\%$,
so the pass the replay simplifies is the one that dominates. The in-kernel adjoint of \cref{sec:method} contributes memory rather
than speed; it is selected by name for that case and does not enter the
ranges above.

\emph{The other baselines.} The PyTorch solvers lag for the separate reason
anticipated in \cref{sec:diff-gpu}: advancing the solve one host-dispatched
operation at a time makes their forward pass alone $27$--$31\x$ slower per
trajectory than a fused kernel, and that overhead, rather than the spread in
step counts, accounts for a substantial part of the \torchodeRange{} gap.
\diffrax{} avoids both the fused kernels' lack of reverse-mode differentiation and the
PyTorch solvers' dispatch overhead, which is why it is the principal baseline of \cref{sec:benchmarks}.

In summary, the gradient is fast because the differentiated computation is a fixed-length replay of recorded steps rather than an adaptive loop differentiated in place: the derivative is the same, but its backward sweep involves no bounded loop and no checkpoint bookkeeping. The controls locate this saving: on the recorded mesh, holding the method fixed, the fixed-step scan computes the gradient \steptoLoopFormGpu{} faster than \diffrax{}'s \code{StepTo} loop on the A100 --- essentially the whole advantage --- while on the CPU the scan holds no such advantage, so the size of the advantage is set by how the GPU executes a fixed scan rather than a checkpointed while-loop. The same saving appears when both engines run in forward mode, where neither stores intermediate values. The in-kernel adjoint adds a memory advantage, and the uneven step
counts across the ensemble contribute only a small factor.

\emph{When to expect the advantage.} The saving comes from removing the
differentiated loop, so it is largest where that loop's overhead dominates the backward sweep. In
the measurements this means state dimensions of up to eight, where a step
costs little relative to the bookkeeping around it; ensembles small enough
that fixed per-step costs still dominate; and workloads that take many
gradients against one recorded mesh, so that the record is paid once. At the
other extreme, as the ensemble grows the margin declines
toward the lower end of \reverseLargeNRange{}, the range measured from
$n = 4096$ to $n = \reverseSweepTop{}$, with the arithmetic dominating only
above roughly $6\times10^{4}$ trajectories. On stiff systems it depends on the accuracy demanded,
from \stiffRangeMax{} at loose accuracy to parity, and slightly below it, at
tight accuracy, where the Rodas5P replay is the right choice. Against \diffrax{}'s forward mode, the cheaper setting when each trajectory carries one to three inputs, the margins are those of \cref{sec:benchmarks-forward-mode}, smaller but of the same shape, and the replay's own forward mode is then the setting to use. For a right-hand side on the general path, one the user has not registered, one the translation of \cref{sec:method-interface} refuses, or one whose state exceeds $d = 64$, the gradient itself costs the same as in these measurements, since the replay is identical. The one-time record then runs as a batched JAX loop on the device, whose time \cref{sec:benchmarks-highorder} reports: a user-defined Lorenz ensemble of \num{8192} trajectories records in \recordDeviceUserLorenz{}.

\section{Discussion and conclusion}\label{sec:discussion}

We introduced \gradsolve{}, an open-source Python library, built on JAX, for
solving and reverse-mode differentiating low-dimensional ODE ensembles on GPUs.
It records the steps an adaptive solver accepts and differentiates a fixed-step
replay of them, so an ensemble solve need not trade a fast integrator for an
inexpensive gradient; the gradient it returns is a discrete adjoint of the same
kind \diffrax{}, the fastest differentiable baseline, produces, obtained from a
simpler computation. The two are complementary: \gradsolve{} specializes this shared adjoint to low-dimensional GPU
ensembles differentiated many times against a recorded mesh, and the library keeps \diffrax{} as both
its principal baseline and a routing fallback. In our benchmarks the reverse-mode gradient runs at
\crossarchRange{} lower cost than \diffrax{}'s checkpointed adjoint
across three GPU generations, and against \diffrax{}'s forward mode, the cheaper
setting for few parameters, the replay keeps a \fwdModeLorenzRange{} speedup in
reverse mode and \fwdReplayLorenzRange{} in forward mode. When no gradient is
required, the forward-only kernel ran \fwdSamePodDouble{} faster than
DiffEqGPU.jl's fused kernel on the low-dimensional ensembles the fused kernels
are built for. The saving carries through to complete parameter recoveries,
which run faster on all three fitting problems, and it does not depend on the
integrator: the seventh-order Verner pair and the fifth-order Rosenbrock method
Rodas5P run on the same path and show a comparable advantage, \mevVernLorenz{}
and \mevRodasRobertson{} per gradient against \diffrax{}
(\cref{sec:benchmarks-highorder}). The controlled experiments of
\cref{sec:whyfast} trace the saving to that simpler form rather than to kernel
fusion or to uneven step counts across the ensemble: on the recorded mesh, with the method held
fixed, the fixed-step scan computes the gradient several times faster than \diffrax{}'s loop on the GPU.

\gradsolve{} is designed for workloads that need reverse-mode
gradients through a GPU ensemble of low-dimensional ODEs. The advantage is
largest on non-stiff systems; it narrows as the ensemble grows on every card
measured (\cref{fig:crossarch}), and on a stiff system at the tightest accuracy
\diffrax{}'s forward mode is the faster choice. The problem itself is
unrestricted: any right-hand side a user writes runs on the general path,
correct and differentiable at any dimension, with a one-time record on the
device and the same per-gradient cost. The fastest, specialized engines apply
within boundaries --- the fused kernels hold each trajectory's state in
registers, which bounds their state dimension (\cref{fig:dimsweep}); the
in-kernel adjoint is verified over a narrower range of state dimensions, and the fused stiff kernel
assumes an autonomous system (\cref{sec:method}) --- and beyond every such
boundary the general path takes over.

In its present form \gradsolve{} solves initial-value problems for ODEs in
the explicit form $\dot y = f(t, y, \theta)$. The fits benchmarked here recover parameters
from final states; losses over many observation times run on the same recorded mesh through the
dense-output replay (\cref{sec:method-interface}) but are not benchmarked, and a criterion for when
a drifting mesh should be re-recorded is left to future work. Stochastic differential
equations, differential-algebraic systems, and event handling are natural
extensions of the construction, left to future work; event handling is the
nearest --- the crossing time found by the adaptive solve can be recorded and
the replay made to land a step on it, with the crossing time's own derivative
supplied separately. Dense output, that is, the state at requested times between the solver's own
steps, comes from the JAX replay engines: by default the replay integrates one extra step
exactly to each requested time, and Rodas5P can instead evaluate its own continuous extension
when that is asked for (\cref{sec:method-interface}); whichever recorder produced the mesh, the
forward-only fused kernels return final states only. Gradients run in double precision on the general path, and in single
precision, with GPU-verified accuracy, for the systems the fused kernels serve.
A user's own right-hand side can be translated into the fused kernels' form
(\cref{sec:method-interface}), and widening that translation is among the
extensions below.

The general path runs wherever JAX runs in double precision: the replay is an
ordinary fixed-step JAX loop (a \code{scan}) that compiles through XLA, JAX's
compiler, and the general recorder is likewise a JAX program. The
fused recorders are Warp kernels, which run on NVIDIA GPUs and, more slowly,
on the CPU. Accelerators without hardware
double precision, such as TPUs and Apple Metal, are outside the measured
setting.

Several of the boundaries above point to concrete extensions. The translation of \cref{sec:method-interface} feeds the Warp kernels; extending it to the hand-written CUDA forward-only kernel, and widening the set of operations it accepts, would bring every user right-hand side to the forward-only speed measured for the built-in fields. A fused kernel that tiles the
state across several threads would raise the dimension limit set by
thread registers without altering the returned gradient. The recorded mesh is
currently reused as-is or rebuilt from scratch; an intermediate policy
refreshing only the trajectories whose parameters have drifted furthest could
benefit long optimization loops.

We release \gradsolve{} as open-source software, built on JAX
\citep{bradbury2018jax}, with fused Warp kernels \citep{nvidia2024warp} for
its recording engines and with \diffrax{} \citep{kidger2021diffrax} available
as both a baseline and a routing fallback, at \url{https://github.com/ECLIPSE-AI4Science/gradsolve}.

\appendix
\renewcommand{\thesection}{\appendixname~\Alph{section}}
\section{Benchmark problem definitions}\label{app:models}

Each benchmark of \cref{sec:benchmarks} is an ensemble
$\dot y = f(t, y, \theta)$, $y \in \mathbb{R}^d$, whose trajectories share a
right-hand side and, where stated below, differ in initial condition or
parameters. For each system we give the vector field, the initial condition,
the spread across the ensemble used by each measurement, and the integration
span. Every measurement is in double precision except where single precision is
stated: the forward comparison against DiffEqGPU.jl
(\cref{sec:benchmarks-forward}) and the single-precision gradient timings of
\cref{sec:benchmarks-reverse}.

\paragraph{Lorenz}
The three-variable attractor \citep{lorenz1963deterministic} evolves as
\begin{align*}
  \dot x &= \sigma\,(y - x), &
  \dot y &= x\,(\rho - z) - y, &
  \dot z &= x\,y - \beta\,z,
\end{align*}
with $\sigma = 10$ and $\beta = 8/3$. The timing benchmarks start every
trajectory at $(1, 0, 0)$ and spread the Rayleigh parameter $\rho$
uniformly over $[0, 21]$ across the ensemble, which spreads per-trajectory
step counts. That range lies below $\rho \approx 24.1$, where the chaotic attractor first
appears, so the timing ensemble exercises transient dynamics. The chaotic
regime enters through the $\rho$-recovery fit below, the galactic orbits,
and a single timing point at $\rho = 28$ (\cref{sec:benchmarks-reverse}). The $\rho$-recovery study of \cref{sec:benchmarks-fitting}
instead starts near $(1, 1, 1)$ with a small ($10^{-2}$) Gaussian
perturbation and spreads $\rho$ across replicates by $\pm 0.15$ decade around
$28$ (roughly $20$ to $40$). Both the timing benchmark and the recovery study
integrate over $t \in [0, 1]$.

\paragraph{Van der Pol}
The relaxation oscillator \citep{vanderpol1926relaxation} reads
\begin{align*}
  \dot x &= v, &
  \dot v &= \mu\,\bigl((1 - x^2)\,v - x\bigr),
\end{align*}
started at $(x, v) = (2, 0)$ with $\mu = 3$; larger $\mu$ stiffens the
oscillator and raises its step count. The reverse-mode sweep of
\cref{sec:benchmarks-reverse} runs every trajectory at $\mu = 3$, so its
ensemble is $n$ copies of one trajectory. Each trajectory is integrated over
$t \in [0, 20]$.

\paragraph{Linear family}
A control problem with a free state dimension $d$: each component grows
independently, $\dot y_i = 1.01\, y_i$, from an initial state drawn uniformly
from $[0, 1)^d$, over $t \in [0, 10]$, with the analytic solution
$y(t) = y_0\, e^{1.01 t}$. Every trajectory needs the same number of steps, and
the right-hand side carries no free parameter. It serves only as the equal-step
control of \cref{tab:dispersion}, as the probe of the fused kernels' dimension
limit (\cref{sec:benchmarks-dim}), and at $d = 32$ in
\cref{sec:benchmarks-highorder}.

\paragraph{Robertson}
The stiff chemical kinetics system \citep{robertson1966solution} reads
\begin{align*}
  \dot y_1 &= -k_1 y_1 + k_3\, y_2 y_3, &
  \dot y_2 &= k_1 y_1 - k_3\, y_2 y_3 - k_2\, y_2^2, &
  \dot y_3 &= k_2\, y_2^2,
\end{align*}
with rates $k_1 = 0.04$, $k_2 = 3\times10^7$, $k_3 = 10^4$ spanning nearly
nine orders of magnitude, started at $(1, 0, 0)$. The per-gradient sweeps of
\cref{sec:benchmarks-stiff} run every trajectory at these rates; the Rodas5P
comparison of \cref{sec:benchmarks-highorder} spreads $k_2$ over
$[3\times10^7,\ 4.5\times10^7]$ across the ensemble; the fitting study
(\cref{sec:benchmarks-fitting}) draws each fit's true $k_1$ and $k_3$ within
$\pm 0.5$ decade of these values and recovers them. Each trajectory is
integrated over $t \in [0, 10^4]$.

\paragraph{HIRES}
The eight-species plant-photochemistry model \citep{schafer1975hires}, a
standard mildly stiff test problem \citep[\S IV.10]{hairer1996solving}, reads
\begingroup\small
\begin{align*}
  \dot y_1 &= -1.71\,y_1 + 0.43\,y_2 + 8.32\,y_3 + 0.0007, &
  \dot y_2 &= 1.71\,y_1 - 8.75\,y_2, \\
  \dot y_3 &= -10.03\,y_3 + 0.43\,y_4 + 0.035\,y_5, &
  \dot y_4 &= 8.32\,y_2 + 1.71\,y_3 - 1.12\,y_4, \\
  \dot y_5 &= -1.745\,y_5 + 0.43\,y_6 + 0.43\,y_7, &
  \dot y_6 &= -280\,y_6 y_8 + 0.69\,y_4 + 1.71\,y_5 - 0.43\,y_6 + 0.69\,y_7, \\
  \dot y_7 &= 280\,y_6 y_8 - 1.81\,y_7, &
  \dot y_8 &= -280\,y_6 y_8 + 1.81\,y_7,
\end{align*}
\endgroup
started at $y = (1, 0, 0, 0, 0, 0, 0, 0.0057)$. HIRES has no free
per-trajectory rate. The Rosenbrock23 comparison of \cref{sec:benchmarks-stiff}
runs every trajectory from this initial condition; the Rodas5P comparison of
\cref{sec:benchmarks-highorder} jitters the nonzero components
multiplicatively by up to $\pm 25\%$, kept non-negative. Each trajectory is integrated from $t = 0$ to $t = 321.8122$, the problem's
standard endpoint.

\paragraph{Galactic logarithmic potential}
Stellar orbits move in a flattened dark-matter halo with the logarithmic
potential \citep{binney2008galactic}
\[
  \Phi(x, y, z) = \tfrac{1}{2}\, v_c^2 \ln\!\bigl(R_c^2 + x^2 + y^2 + z^2/q^2\bigr).
\]
The state is the six-dimensional phase-space vector
$(x, y, z, v_x, v_y, v_z)$ with $\dot{\mathbf r} = \mathbf v$,
$\dot{\mathbf v} = -\nabla\Phi$, i.e.
\[
  \dot v_x = -\frac{v_c^2\, x}{D}, \quad
  \dot v_y = -\frac{v_c^2\, y}{D}, \quad
  \dot v_z = -\frac{v_c^2\, z}{q^2\, D}, \qquad
  D = R_c^2 + x^2 + y^2 + z^2/q^2,
\]
in scaled units with $v_c = 1$, $R_c = 0.2$, flattening $q = 0.7$. For
$q < 1$ many of these orbits are chaotic; $q$ is the parameter recovered in
\cref{sec:benchmarks-fitting}. All trajectories share the halo parameters,
and the ensemble varies the stellar initial conditions: launch radii
$r \in [0.3, 1.2]$ and azimuths $\phi \in [0, 2\pi)$, heights $z \in [-0.4, 0.4]$,
tangential speeds $0.7$--$1.0$ times $v_c$, and vertical
velocities $v_z \in [-0.3, 0.3]$, spanning loop orbits, which circulate around
the center, and box orbits, which oscillate through it. Each trajectory is integrated over $t \in [0, 6]$.

\paragraph{Robertson neural right-hand side}
A stiff neural differential equation
\citep{rackauckas2020universal, chen2018neuralode} is built from Robertson by
replacing the autocatalytic term $k_2\, y_2^2$ with a small neural network
$\mathrm{NN}_\theta : \mathbb{R}^3 \to \mathbb{R}$ in both the sink and
source terms, so mass is conserved:
\begin{align*}
  \dot y_1 &= -k_1 y_1 + k_3\, y_2 y_3, &
  \dot y_2 &= k_1 y_1 - k_3\, y_2 y_3 - \mathrm{NN}_\theta(y), &
  \dot y_3 &= \mathrm{NN}_\theta(y),
\end{align*}
with $k_1 = 0.04$, $k_3 = 10^4$ and Robertson's initial condition and integration span.
The network is a multilayer perceptron with input dimension $3$, output
dimension $1$, two hidden layers of width $16$, $\tanh$ hidden activations, and
a softplus output, keeping the replaced term non-negative. The parameters $\theta$ are the network
weights; the gradient of the fitting loss with respect to $\theta$ flows
through the stiff solve. The batch-size sweep of \cref{fig:udebscaling}
spreads $k_1$ per trajectory over $[k_1, 1.5\,k_1]$.

\paragraph{Lorenz-96}
Used only for the high-dimensional gradient check of \ref{app:gradchecks}, not
as a timing benchmark, the cyclic Lorenz-96 system has right-hand side
\[
  \dot y_i = (y_{i+1} - y_{i-2})\,y_{i-1} - y_i + F,
\]
with the indices $i$ taken cyclically over the $d$ state components and a
constant forcing $F$; the check runs at $d = 96$.
\section{Benchmark protocol and reproducibility}\label{app:repro}

This appendix records how the measurements of \cref{sec:benchmarks} were made, in
enough detail to repeat them: the rule by which accuracy is matched
(\ref{app:protocol-matching}), the configuration of every baseline
(\ref{app:protocol-baseline}), the timing rule and its uncertainties
(\ref{app:protocol-timing}), the protocol of the fitting studies
(\ref{app:protocol-fitting}), the hardware and software (\ref{app:versions}), and
the checks that the returned gradients are correct (\ref{app:gradchecks}).

\subsection{Matching rule}\label{app:protocol-matching}

Solvers of different order reach different accuracy at the same nominal tolerance,
so ratios between methods of different order are read at matched achieved accuracy.
\gradsolve{} runs first and its error against a high-accuracy reference is recorded:
the median, over all components of all final states in the ensemble, of the
componentwise relative error $\lvert \hat y - y_{\mathrm{ref}}\rvert /
\max(\lvert y_{\mathrm{ref}}\rvert, 10^{-30})$, with $y_{\mathrm{ref}}$ a
high-accuracy reference --- analytic where one exists, otherwise \code{scipy}
Radau for stiff systems and DOP853 for the rest, at $\tau_{\mathrm{rel}} = 10^{-11}$,
$\tau_{\mathrm{abs}} = 10^{-12}$ in double precision. The median is the matched
quantity; \ref{app:gradchecks} also reports the maximum and, where a system's
components differ in scale, the per-species error. The baseline's relative tolerance is then bisected,
with $\tau_{\mathrm{abs}} = 10^{-3}\,\tau_{\mathrm{rel}}$ throughout, until its own
achieved error lies within $25\%$ of ours. Among the tolerances inside that band,
the one whose error is not smaller than ours is quoted, so the baseline is never
credited with accuracy it was not asked for; and no match is taken below
$\tau_{\mathrm{rel}} = 10^{-10}$. Accuracy is matched on the solution: each
engine's gradient is exact for its own discretization, and the two gradients are
not separately matched. Where a range is read off a work--precision sweep, the
baseline's cost at \gradsolve{}'s achieved error is obtained by log--log
interpolation between the baseline's measured points, within the overlap of the two
curves' achieved-error ranges only; no ratio is extrapolated.

\subsection{Baseline configuration}\label{app:protocol-baseline}

Every device is synchronized before a timestamp is read:
\code{block\_until\_ready} in JAX, and \code{torch.cuda.synchronize} in PyTorch.

The \diffrax{} baseline is \code{diffeqsolve} with \code{Tsit5} for the non-stiff
systems and \code{Kvaerno5} for the stiff ones (\cref{sec:benchmarks-stiff}); the
adaptive \code{PIDController} at its default coefficients, which reduce to an
integral controller as in \gradsolve{}'s recorders; the
default \code{RecursiveCheckpointAdjoint}; \code{dt0=None}, so that the controller
chooses the initial step; \code{SaveAt(t1=True)}; and \code{throw=False}; mapped
over the ensemble with \code{vmap}. Both engines are compiled by one outer
\code{jit} around \code{grad}, with no further \code{jit} placed inside it
(\code{diffeqsolve} carries its own). The stiff baseline solves its implicit stages
with a full Newton iteration (\code{optimistix.Newton} \citep{rader2024optimistix},
whose linear solves use lineax \citep{rader2023lineax}, with the controller's tolerances) in place of the default chord iteration
(\code{VeryChord}), which reuses one Jacobian factorization per step. On the
Robertson system at matched achieved accuracy, measured on a CPU at $n = 64$, the chord
setting rejects about half of its attempted steps, accepts three to four times as
many steps as the Newton setting, and takes $3.6$--$3.9\x$ longer per gradient; the
Newton setting is therefore the faster one and is the one timed. This comparison was made on a CPU; the GPU runs use the same Newton setting, whose lower step count is a property of the numerics rather than of the device.

The step limit \code{max\_steps} is sized from a preliminary solve of the same
ensemble. The checkpoint count is the fastest of three measured settings: one
checkpoint per attempted step, with the step limit tightened to the attempted count
so that nothing is recomputed (the solve is re-run at that limit to confirm that no
trajectory reaches it); one and a half times the attempted count on the sized step
limit; and the library default, which derives the count from the step limit. The
choice and its margin over the runner-up are recorded for each measurement, that is,
for each combination of problem, ensemble size and pair of engines. In the fitting
loops the parameters move away from the point where that step limit was sized.
There \diffrax{} instead runs one checkpoint per step of a step limit sized with
headroom at the starting point, so it never recomputes and cannot exhaust the limit
silently; that setting was within \fitPinnedMargin{} of the fastest measured one,
and at the end of every fit the \diffrax{} solve is re-run at the final parameters
to confirm that no trajectory reached the step limit (its tolerance, matched at the
starting point, is not re-matched). The library's own built-in \diffrax{} engine
(\cref{sec:method}) uses \diffrax{}'s defaults and is not the configuration timed
here.

All baselines are timed in reverse mode, the mode the neural right-hand side of
\cref{sec:benchmarks-fitting} requires. For the systems with one to three inputs
per trajectory, \diffrax{}'s forward-mode option (\code{ForwardMode}) was timed as
well, in a separate A100 session, with the same solver, controller, tolerances and
step limit as the reverse arm and the checkpointed adjoint replaced by
\code{ForwardMode}.

Because the loss is a sum over trajectories and no trajectory depends on another's inputs, the gradient is assembled from one forward pass per input, each pass changing that input in every trajectory at once. Two placements were timed, one pass over the whole ensemble and \code{jacfwd} (forward-mode Jacobian) applied per trajectory inside the \code{vmap}, and the faster of the two is quoted. Both return the reverse arm's gradient to within $10^{-13}$, and both solve the same trajectories as the reverse arm to roundoff. Pushing one pass per trajectory and input through the whole ensemble, which costs $n$ times more, was not timed.

The replay was differentiated in forward mode by the same
rule. The two reverse arms were re-measured in that session and reproduce the
ranges quoted in \cref{sec:benchmarks} within the run-to-run spread. \diffrax{} also offers a continuous adjoint (\code{BacksolveAdjoint}), which its
documentation does not recommend because the gradient is approximate; it is not
timed.

The two PyTorch solvers were timed on a second A100 machine (PCIe), in a separate
process from the JAX measurements, because PyTorch's bundled CUDA libraries
otherwise interfere with JAX; the ratios against them therefore compare across two
A100 machines, which differ by about 18\% at the smallest ensemble size, above the run-to-run spread on one machine. torchode is timed under \code{torch.compile}, its intended fast configuration, and torchdiffeq's continuous adjoint eagerly. In the fixed-mesh control of
\cref{sec:whyfast}, a uniform Robertson mesh of our step count failed on every
trajectory; the number reported comes from a geometrically graded mesh, which
supplies a mesh shape as well as replacing the controller.

\subsection{Timing and uncertainty}\label{app:protocol-timing}

Each per-gradient wall time is the best of three timed calls (five for the
higher-order engines of \cref{sec:benchmarks-highorder}) and each fitting-loop time
the mean over repetitions, both after a warm-up pass that absorbs one-time
compilation and, for the replay engines, the one-time recording; the run-to-run
spread of such a measurement is quoted in \cref{sec:benchmarks}. The best-of-N
choice isolates the computation from transient scheduling and contention on a
shared host, the usual convention for timing a compiled kernel; the fitting-loop
times run long enough to average such transients and are reported as means, with
\cref{fig:fitting} drawing their standard deviation as error bars. The
per-repetition timings behind the reverse figures are kept in the released data,
so any other statistic can be formed from them; on the log-scaled reverse figures
the \timingSpread{} spread is smaller than the plotted markers, so it is reported
here rather than drawn. That spread is comparable to the narrowest margins ---
the near-parity stiff points at tight accuracy, which should be read with it in
mind --- while the large reverse-mode margins sit far above it.
Two sources of uncertainty act on the ratios of \cref{sec:benchmarks-highorder} in
opposite directions. The matching procedure lowers \diffrax{}'s tolerance by
bisection until its achieved error reaches ours, never the reverse, so each
baseline ended slightly less accurate than \gradsolve{} (achieved-error ratios
1.01--1.12) and takes fewer steps than an exact match would, which understates the
ratios. Against that, a finer ten-point sweep of \diffrax{}'s checkpoint setting
found settings up to $2\%$ faster than the three settings measured, which
overstates each ratio by at most that much.

\subsection{Fitting protocol}\label{app:protocol-fitting}

The fitting studies of \cref{sec:benchmarks-fitting} run $B$ concurrent fits,
$B \in \{1, 8, 64, 256\}$, each recovering its own parameter, with both engines
under the identical optimizer. The neural network study of \cref{fig:udebscaling}
reuses the symbol $B$ for its minibatch size, the number of trajectories per
training step. The Lorenz and galactic fits run 400 optimizer
updates; the Robertson fit stops once its loss falls below a fixed tolerance. The
recorded mesh is treated as data for the duration of a fit: no reported fit
re-records midway, and re-recording at a fixed cadence, when used, is timed inside
the loop. Any combination of problem and $B$ where either engine failed to recover
its parameter is dropped, since a failed baseline is not evidence of a speedup.
The optimizer is Adam \citep{kingma2015adam} at a learning rate of $0.05$ (Lorenz),
$0.02$ (galactic) or $0.1$ (Robertson); each fit starts a tenth of a decade
(Lorenz), a quarter decade (galactic) or a full decade (Robertson) from the true
value and recovers it from final states perturbed by relative Gaussian noise of
magnitude $10^{-3}$; the Robertson loss tolerance is $10^{-5}$, and all runs use
seed $0$, a single data realization, and each fit's optimizer-loop time is the mean over two timed repetitions. Recovery is declared
when the parameter returns to within a relative $0.02$ (Lorenz), $0.03$ (galactic)
or $0.05$ in $\log_{10}$ (Robertson) of its true value. Both engines recover it to
the same accuracy: on Lorenz the final $\log_{10}$ error runs from \num{7e-5} at
$B = 1$ to \num{2.3e-3}, on galactic the relative error from \num{2.9e-4} down to
\num{7.2e-6} as $B$ grows, and on Robertson the $\log_{10}$ error stays below
\num{0.02} for $B \le 64$. The one dropped combination is the stiff Robertson fit
at $B = 256$, where neither engine reached the loss tolerance within its step cap
($\log_{10}$ error $\approx 1.13$ for both), the absent point in \cref{fig:fitting}.

\subsection{Hardware and software}\label{app:versions}

Most experiments run in double precision on a single GPU: an NVIDIA A100
(80\,GB), an H100, or a consumer RTX 4090 (24\,GB, whose double-precision
arithmetic rate is one sixty-fourth of its single-precision rate). The A100 comes
in two packagings, PCIe and SXM4, which are named where they differ, and the A100
measurements were made on two machines. The CPU experiment of \cref{sec:whyfast}
was run on an \cpuLaptop{} and the CPU curve of \cref{fig:reverse}(b) on the CPU of
the machine hosting the A100. No two frameworks ever share a device within one
measurement. The A100 sweeps ran JAX~0.10.2 with \diffrax{}~0.7.2 and Warp~1.15;
the DiffEqGPU.jl comparison ran Julia~1.12.5 with DiffEqGPU.jl~3.15.2 and
CUDA.jl~6. \Cref{tab:versions} lists the software each measurement ran under.

\begin{table}[htb]
  \centering\footnotesize
  \caption{Software environment of the measurements reported here.
  The A100 measurements were made on two machines, hence two driver versions. The Julia
  row gives the forward comparison of \cref{sec:benchmarks-forward} first and
  the reverse-mode attempt of \cref{sec:benchmarks-reverse} second.}
  \label{tab:versions}
  \begin{tabular}{ll}
    \toprule
    Component & Version \\
    \midrule
    GPUs and drivers & A100-SXM4-80GB (580.126.16, 570.172.08), \\
                     & H100 80GB HBM3 (580.126.09), RTX 4090 (580.126.09) \\
    CUDA             & 12.4 (JAX build) \\
    JAX / \diffrax{} / equinox / optimistix / lineax & 0.10.2 / 0.7.2 / 0.13.8 / 0.1.0 / 0.1.1 \\
    Warp             & 1.15.0 \\
    Host CPUs        & \cpuPodHost{} (first A100 machine), \\
                     & \cpuPodHostB{} (second A100 machine), \\
                     & Intel Xeon Platinum 8480+ (H100 machine), \\
                     & AMD EPYC 7K62 48-Core (RTX 4090 machine), \\
                     & \cpuLaptop{} (CPU experiment of \cref{sec:whyfast}) \\
    PyTorch / torchode (\code{torch.compile} measurement) & 2.4.1+cu124 / 1.0.1 \\
    Julia / DiffEqGPU.jl / CUDA.jl & 1.12.5 / 3.15.2 / 6; \; 1.12.6 / 3.15.3 / 6.2.1 \\
    \bottomrule
  \end{tabular}
\end{table}

\subsection{Gradient checks}\label{app:gradchecks}

The exactness of the gradient (\cref{sec:method}) is a property of the
differentiated program rather than of an accelerator, so a finite-difference test
on any machine can verify it. The check builds a small ensemble, differentiates the benchmark loss
$\ell = \lVert y(t_1)\rVert^2$ summed over the ensemble through it with \gradsolve{},
and compares the result to a central finite-difference gradient of the same loss
(relative step $10^{-6}$ along a seeded unit direction, ensemble seed $0$),
perturbing one trajectory's parameter inside the same recorded replay (the
convention of \cref{sec:method}). On a laptop
CPU the two agree to a relative difference of about \num{9e-11}; on the A100 the
stiff Robertson replay agrees with the finite-difference estimate to
\fdNoiseFloor{}. Where the finite-difference check is itself uninformative --- for
the linear family at $d = 32$, whose rescaling loss drives the central-difference
signal to the roundoff floor --- the replay's gradient is compared instead against
the system's closed-form sensitivity, and agrees with it to \num{1.6e-7} relative,
the order of the method's forward error (\cref{sec:benchmarks-highorder}).

The checks so far perturb inputs \emph{inside} the recorded replay, so they confirm
that automatic differentiation of the replay is correct; they do not by themselves
show that this gradient is an accurate sensitivity of the underlying system. A
separate test does. For a non-stiff system (Lorenz) and a stiff one (Robertson) we
compare the record-and-replay gradient of $\lVert y(t_1)\rVert^2$ against an
\emph{independent} high-accuracy sensitivity: central finite differences of a
separately, tightly solved reference (the accuracy-matching reference of
\ref{app:protocol-matching}, not the replay), swept over the step size to expose
the finite-difference plateau before roundoff. The two agree to \gradAccLorenz{}
on Lorenz and \gradAccRobertson{} on Robertson, the level of each system's own
forward discretization error rather than machine precision, so the replay returns
an accurate sensitivity and not merely a self-consistent derivative
(\cref{tab:gradacc}).
A third check compares the two engines head to head, on a shared mesh. Running
\diffrax{}'s Tsit5 on \gradsolve{}'s own recorded mesh through its \code{StepTo} controller
--- the same method stepping to the same recorded times --- its gradient of
$\lVert y(t_1)\rVert^2$ agrees with the replay's to \sharedMeshLorenz{} on Lorenz and
\sharedMeshVdp{} on Van der Pol, at machine precision. This confirms directly what
\cref{sec:method-construction} argues: on the same method and the same mesh the two return the same
discrete adjoint, not two gradients that each merely match finite differences. The forward-state error is reported there as a median and a
maximum; on Robertson, whose species span decades, the maximum (\errMaxRobertson{})
sits close to the median (\errMedRobertson{}), and the per-component maxima are
\num{3.3e-5}, \num{4.4e-5} and \num{3.6e-6}, so the small middle species is resolved as
well as the others, so the pooled median is
representative rather than masking a poorly resolved species.

\begin{table}[tb]
  \centering\footnotesize
  \caption{Gradient accuracy against an independent high-accuracy sensitivity (CPU,
  double precision). For each system the record-and-replay gradient of
  $\lVert y(t_1)\rVert^2$ is compared with central finite differences of a
  separately, tightly solved reference (\ref{app:protocol-matching}), reported as the
  relative difference of the two gradients; the achieved forward-state error is given
  as the median and the maximum over the ensemble.}
  \label{tab:gradacc}
  \begin{tabular}{lccc}
    \toprule
    System & Gradient vs independent sensitivity & Forward error (median) & Forward error (max) \\
    \midrule
    Lorenz (non-stiff)   & \gradAccLorenz{}    & \errMedLorenz{}    & \errMaxLorenz{} \\
    Robertson (stiff)    & \gradAccRobertson{} & \errMedRobertson{} & \errMaxRobertson{} \\
    \bottomrule
  \end{tabular}
\end{table}

On the non-stiff side, two independent implementations of the
backward sweep, the JAX replay and a Warp kernel differentiated in place, agree to
about \num{3.5e-11} at state dimension \num{96} on the Lorenz-96 system
(\ref{app:models}); the
finite-difference check of that kernel was made through $d = 32$, and its stiff
counterpart, the in-kernel adjoint, through $d \approx 12$. Inside the in-kernel
adjoint two pieces are written by hand rather than generated. The kernel solves its
linear systems in place, overwriting the numbers it reads, and Warp's automatic
differentiation does not propagate correctly through such overwrites, so the
transpose-solve rule of the backward sweep is written out explicitly. Its
Jacobian comes from a formula registered with the library, hand-written for the
built-in fields and derived by automatic differentiation for a translated user
field (\cref{sec:method-interface}), rather than from differentiating the
right-hand side inside the kernel. A separate run of the same comparison with the fused recorder in place of the general recorder returns the same mesh to roundoff and hence the same gradient, the observation \cref{sec:whyfast} relies on. The general recorder's compiled loop and a plain Python loop over the same trial steps agree exactly in accepted and rejected step counts, with step sizes agreeing to about \num{1e-8} relative (the compiled loop uses fused multiply-add arithmetic).

\section*{Acknowledgments}
ASM thanks Patrick Kidger for helpful comments on the manuscript, and acknowledges funding
from a Leverhulme Trust Research Leadership Award.

\section*{Data and code availability}
The \gradsolve{} code, together with notebooks that re-measure the reverse-mode comparison
against \diffrax{} and the forward-only comparison against DiffEqGPU.jl on the reader's own
hardware, is available at \url{https://github.com/ECLIPSE-AI4Science/gradsolve}.

\bibliographystyle{plainnat}
\bibliography{refs}

\begin{thebibliography}{58}
\providecommand{\natexlab}[1]{#1}
\providecommand{\url}[1]{\texttt{#1}}
\expandafter\ifx\csname urlstyle\endcsname\relax
  \providecommand{\doi}[1]{doi: #1}\else
  \providecommand{\doi}{doi: \begingroup \urlstyle{rm}\Url}\fi

\bibitem[Alexe and Sandu(2009)]{alexe2009forward}
Mihai Alexe and Adrian Sandu.
\newblock Forward and adjoint sensitivity analysis with continuous explicit
  {Runge--Kutta} schemes.
\newblock \emph{Applied Mathematics and Computation}, 208\penalty0
  (2):\penalty0 328--346, 2009.
\newblock \doi{10.1016/j.amc.2008.11.035}.

\bibitem[Baydin et~al.(2018)Baydin, Pearlmutter, Radul, and
  Siskind]{baydin2018automatic}
At{\i}l{\i}m~G{\"u}ne{\c s} Baydin, Barak~A. Pearlmutter, Alexey~Andreyevich
  Radul, and Jeffrey~Mark Siskind.
\newblock Automatic differentiation in machine learning: a survey.
\newblock \emph{Journal of Machine Learning Research}, 18\penalty0
  (153):\penalty0 1--43, 2018.
\newblock \doi{10.48550/arXiv.1502.05767}.
\newblock {arXiv}:1502.05767.

\bibitem[Betts(2010)]{betts2010practical}
John~T. Betts.
\newblock \emph{Practical Methods for Optimal Control and Estimation Using
  Nonlinear Programming}.
\newblock Society for Industrial and Applied Mathematics, Philadelphia, second
  edition, 2010.
\newblock \doi{10.1137/1.9780898718577}.

\bibitem[Binney and Tremaine(2008)]{binney2008galactic}
James Binney and Scott Tremaine.
\newblock \emph{Galactic Dynamics}.
\newblock Princeton University Press, 2nd edition, 2008.
\newblock ISBN 978-0-691-13026-2.
\newblock \doi{10.1515/9781400828722}.

\bibitem[Bradbury et~al.(2018)Bradbury, Frostig, Hawkins, Johnson, Katariya,
  Leary, Maclaurin, Necula, Paszke, Vander{P}las, Wanderman-{M}ilne, and
  Zhang]{bradbury2018jax}
James Bradbury, Roy Frostig, Peter Hawkins, Matthew~James Johnson, Yash
  Katariya, Chris Leary, Dougal Maclaurin, George Necula, Adam Paszke, Jake
  Vander{P}las, Skye Wanderman-{M}ilne, and Qiao Zhang.
\newblock {JAX}: composable transformations of {P}ython+{N}um{P}y programs.
\newblock Software, 2018.
\newblock \url{http://github.com/jax-ml/jax}.

\bibitem[Cao et~al.(2003)Cao, Li, Petzold, and Serban]{cao2003adjoint}
Yang Cao, Shengtai Li, Linda Petzold, and Radu Serban.
\newblock Adjoint sensitivity analysis for differential-algebraic equations:
  The adjoint {DAE} system and its numerical solution.
\newblock \emph{SIAM Journal on Scientific Computing}, 24\penalty0
  (3):\penalty0 1076--1089, 2003.
\newblock \doi{10.1137/S1064827501380630}.

\bibitem[Chen(2018)]{chen2018torchdiffeq}
Ricky T.~Q. Chen.
\newblock {torchdiffeq}: differentiable {ODE} solvers with full {GPU} support
  and {O(1)}-memory backpropagation.
\newblock Software, 2018.
\newblock \url{https://github.com/rtqichen/torchdiffeq}.

\bibitem[Chen et~al.(2018)Chen, Rubanova, Bettencourt, and
  Duvenaud]{chen2018neuralode}
Ricky T.~Q. Chen, Yulia Rubanova, Jesse Bettencourt, and David Duvenaud.
\newblock Neural ordinary differential equations.
\newblock In \emph{Advances in Neural Information Processing Systems
  (NeurIPS)}, 2018.
\newblock \doi{10.48550/arXiv.1806.07366}.
\newblock {arXiv}:1806.07366.

\bibitem[Dormand and Prince(1980)]{dormand1980family}
J.~R. Dormand and P.~J. Prince.
\newblock A family of embedded {Runge--Kutta} formulae.
\newblock \emph{Journal of Computational and Applied Mathematics}, 6\penalty0
  (1):\penalty0 19--26, 1980.
\newblock \doi{10.1016/0771-050X(80)90013-3}.

\bibitem[Farrell et~al.(2013)Farrell, Ham, Funke, and
  Rognes]{farrell2013automated}
Patrick~E. Farrell, David~A. Ham, Simon~W. Funke, and Marie~E. Rognes.
\newblock Automated derivation of the adjoint of high-level transient finite
  element programs.
\newblock \emph{SIAM Journal on Scientific Computing}, 35\penalty0
  (4):\penalty0 C369--C393, 2013.
\newblock \doi{10.1137/120873558}.

\bibitem[Gholaminejad et~al.(2019)Gholaminejad, Keutzer, and
  Biros]{gholami2019anode}
Amir Gholaminejad, Kurt Keutzer, and George Biros.
\newblock {ANODE}: Unconditionally accurate memory-efficient gradients for
  neural {ODEs}.
\newblock In \emph{Proceedings of the Twenty-Eighth International Joint
  Conference on Artificial Intelligence (IJCAI)}, pages 730--736, 2019.
\newblock \doi{10.24963/ijcai.2019/103}.

\bibitem[Giles and Pierce(2000)]{giles2000adjoint}
Michael~B. Giles and Niles~A. Pierce.
\newblock An introduction to the adjoint approach to design.
\newblock \emph{Flow, Turbulence and Combustion}, 65\penalty0 (3-4):\penalty0
  393--415, 2000.
\newblock \doi{10.1023/A:1011430410075}.

\bibitem[Griewank and Walther(2000)]{griewank2000revolve}
Andreas Griewank and Andrea Walther.
\newblock Algorithm 799: revolve: an implementation of checkpointing for the
  reverse or adjoint mode of computational differentiation.
\newblock \emph{ACM Transactions on Mathematical Software}, 26\penalty0
  (1):\penalty0 19--45, 2000.
\newblock \doi{10.1145/347837.347846}.

\bibitem[Griewank and Walther(2008)]{griewank2008evaluating}
Andreas Griewank and Andrea Walther.
\newblock \emph{Evaluating Derivatives: Principles and Techniques of
  Algorithmic Differentiation}.
\newblock Society for Industrial and Applied Mathematics, 2nd edition, 2008.
\newblock \doi{10.1137/1.9780898717761}.

\bibitem[Hager(2000)]{hager2000runge}
William~W. Hager.
\newblock Runge--kutta methods in optimal control and the transformed adjoint
  system.
\newblock \emph{Numerische Mathematik}, 87\penalty0 (2):\penalty0 247--282,
  2000.
\newblock \doi{10.1007/s002110000178}.

\bibitem[Hairer and Wanner(1996)]{hairer1996solving}
Ernst Hairer and Gerhard Wanner.
\newblock \emph{Solving Ordinary Differential Equations II: Stiff and
  Differential-Algebraic Problems}.
\newblock Springer, 1996.
\newblock \doi{10.1007/978-3-642-05221-7}.

\bibitem[Hairer et~al.(1993)Hairer, N{\o}rsett, and Wanner]{hairer1993solving}
Ernst Hairer, Syvert~P. N{\o}rsett, and Gerhard Wanner.
\newblock \emph{Solving Ordinary Differential Equations I: Nonstiff Problems}.
\newblock Springer, 2nd rev. edition, 1993.
\newblock \doi{10.1007/978-3-540-78862-1}.

\bibitem[Hindmarsh et~al.(2005)Hindmarsh, Brown, Grant, Lee, Serban, Shumaker,
  and Woodward]{hindmarsh2005sundials}
Alan~C. Hindmarsh, Peter~N. Brown, Keith~E. Grant, Steven~L. Lee, Radu Serban,
  Dan~E. Shumaker, and Carol~S. Woodward.
\newblock {SUNDIALS}: Suite of nonlinear and differential/algebraic equation
  solvers.
\newblock \emph{ACM Transactions on Mathematical Software}, 31\penalty0
  (3):\penalty0 363--396, 2005.
\newblock \doi{10.1145/1089014.1089020}.
\newblock \url{https://github.com/LLNL/sundials}.

\bibitem[Kidger(2021{\natexlab{a}})]{kidger2021diffrax}
Patrick Kidger.
\newblock {Diffrax}: numerical differential equation solvers in {JAX}.
\newblock Software, 2021{\natexlab{a}}.
\newblock \url{https://github.com/patrick-kidger/diffrax}.

\bibitem[Kidger(2021{\natexlab{b}})]{kidger2022neural}
Patrick Kidger.
\newblock \emph{On Neural Differential Equations}.
\newblock PhD thesis, University of Oxford, 2021{\natexlab{b}}.
\newblock {arXiv}:2202.02435.

\bibitem[Kidger and Garcia(2021)]{kidger2021equinox}
Patrick Kidger and Cristian Garcia.
\newblock Equinox: neural networks in {JAX} via callable {PyTrees} and filtered
  transformations, 2021.
\newblock {arXiv}:2111.00254, \url{https://github.com/patrick-kidger/equinox}.

\bibitem[Kingma and Ba(2015)]{kingma2015adam}
Diederik~P. Kingma and Jimmy Ba.
\newblock Adam: A method for stochastic optimization.
\newblock In \emph{International Conference on Learning Representations
  (ICLR)}, 2015.
\newblock \doi{10.48550/arXiv.1412.6980}.
\newblock {arXiv}:1412.6980.

\bibitem[Kv{\ae}rn{\o}(2004)]{kvaerno2004singly}
Anne Kv{\ae}rn{\o}.
\newblock Singly diagonally implicit {Runge--Kutta} methods with an explicit
  first stage.
\newblock \emph{BIT Numerical Mathematics}, 44\penalty0 (3):\penalty0 489--502,
  2004.
\newblock \doi{10.1023/B:BITN.0000046811.70614.38}.

\bibitem[Lienen and G{\"u}nnemann(2022)]{lienen2022torchode}
Marten Lienen and Stephan G{\"u}nnemann.
\newblock torchode: A parallel {ODE} solver for {PyTorch}.
\newblock In \emph{The Symbiosis of Deep Learning and Differential Equations
  II, NeurIPS Workshop}, 2022.
\newblock \doi{10.48550/arXiv.2210.12375}.
\newblock {arXiv}:2210.12375, \url{https://github.com/martenlienen/torchode}.

\bibitem[Lindholm et~al.(2008)Lindholm, Nickolls, Oberman, and
  Montrym]{lindholm2008nvidia}
Erik Lindholm, John Nickolls, Stuart Oberman, and John Montrym.
\newblock {NVIDIA Tesla}: A unified graphics and computing architecture.
\newblock \emph{IEEE Micro}, 28\penalty0 (2):\penalty0 39--55, 2008.
\newblock \doi{10.1109/MM.2008.31}.

\bibitem[Lorenz(1963)]{lorenz1963deterministic}
Edward~N. Lorenz.
\newblock Deterministic nonperiodic flow.
\newblock \emph{Journal of the Atmospheric Sciences}, 20\penalty0 (2):\penalty0
  130--141, 1963.
\newblock \doi{10.1175/1520-0469(1963)020<0130:DNF>2.0.CO;2}.

\bibitem[Ma et~al.(2021)Ma, Dixit, Innes, Guo, and
  Rackauckas]{ma2021comparison}
Yingbo Ma, Vaibhav Dixit, Michael~J. Innes, Xingjian Guo, and Christopher
  Rackauckas.
\newblock A comparison of automatic differentiation and continuous sensitivity
  analysis for derivatives of differential equation solutions.
\newblock In \emph{2021 IEEE High Performance Extreme Computing Conference
  (HPEC)}, pages 1--9, 2021.
\newblock \doi{10.1109/HPEC49654.2021.9622796}.
\newblock {arXiv}:1812.01892.

\bibitem[Nagy et~al.(2022)Nagy, Plavecz, and Heged{\H{u}}s]{nagy2022mpgos}
D{\'a}niel Nagy, Lambert Plavecz, and Ferenc Heged{\H{u}}s.
\newblock The art of solving a large number of non-stiff, low-dimensional
  ordinary differential equation systems on {GPUs} and {CPUs}.
\newblock \emph{Communications in Nonlinear Science and Numerical Simulation},
  112:\penalty0 106521, 2022.
\newblock \doi{10.1016/j.cnsns.2022.106521}.
\newblock {MPGOS},
  \url{https://github.com/FerencHegedus/Massively-Parallel-GPU-ODE-Solver}.

\bibitem[{NVIDIA}(2024)]{nvidia2024warp}
{NVIDIA}.
\newblock {NVIDIA Warp}: A python framework for high-performance gpu simulation
  and graphics.
\newblock \url{https://github.com/NVIDIA/warp}, 2024.

\bibitem[{NVIDIA Corporation}(2025)]{nvidia2025cuda}
{NVIDIA Corporation}.
\newblock {CUDA C++ Programming Guide}.
\newblock \url{https://docs.nvidia.com/cuda/cuda-c-programming-guide/}, 2025.
\newblock Release 12.x; the per-thread register limit of 255 is given in the
  compute-capability table.

\bibitem[Onken and Ruthotto(2020)]{onken2020discretize}
Derek Onken and Lars Ruthotto.
\newblock Discretize-optimize vs. optimize-discretize for time-series
  regression and continuous normalizing flows, 2020.
\newblock {arXiv}:2005.13420.

\bibitem[Paszke et~al.(2019)Paszke, Gross, Massa, Lerer, Bradbury, Chanan,
  Killeen, Lin, Gimelshein, Antiga, Desmaison, K{\"o}pf, Yang, DeVito, Raison,
  Tejani, Chilamkurthy, Steiner, Fang, Bai, and Chintala]{paszke2019pytorch}
Adam Paszke, Sam Gross, Francisco Massa, Adam Lerer, James Bradbury, Gregory
  Chanan, Trevor Killeen, Zeming Lin, Natalia Gimelshein, Luca Antiga, Alban
  Desmaison, Andreas K{\"o}pf, Edward Yang, Zachary DeVito, Martin Raison,
  Alykhan Tejani, Sasank Chilamkurthy, Benoit Steiner, Lu~Fang, Junjie Bai, and
  Soumith Chintala.
\newblock {PyTorch}: An imperative style, high-performance deep learning
  library.
\newblock In \emph{Advances in Neural Information Processing Systems
  (NeurIPS)}, 2019.
\newblock \doi{10.48550/arXiv.1912.01703}.
\newblock {arXiv}:1912.01703, \url{https://github.com/pytorch/pytorch}.

\bibitem[Rackauckas and Nie(2017)]{rackauckas2017differentialequations}
Christopher Rackauckas and Qing Nie.
\newblock {DifferentialEquations.jl} -- a performant and feature-rich ecosystem
  for solving differential equations in {Julia}.
\newblock \emph{Journal of Open Research Software}, 5\penalty0 (1):\penalty0
  15, 2017.
\newblock \doi{10.5334/jors.151}.
\newblock \url{https://github.com/SciML/DifferentialEquations.jl}.

\bibitem[Rackauckas et~al.(2020)Rackauckas, Ma, Martensen, Warner, Zubov,
  Supekar, Skinner, Ramadhan, and Edelman]{rackauckas2020universal}
Christopher Rackauckas, Yingbo Ma, Julius Martensen, Collin Warner, Kirill
  Zubov, Rohit Supekar, Dominic Skinner, Ali Ramadhan, and Alan Edelman.
\newblock Universal differential equations for scientific machine learning,
  2020.
\newblock {arXiv}:2001.04385.

\bibitem[Rader et~al.(2023)Rader, Lyons, and Kidger]{rader2023lineax}
Jason Rader, Terry Lyons, and Patrick Kidger.
\newblock Lineax: unified linear solves and linear least-squares in {JAX} and
  {Equinox}, 2023.
\newblock {arXiv}:2311.17283, \url{https://github.com/patrick-kidger/lineax}.

\bibitem[Rader et~al.(2024)Rader, Lyons, and Kidger]{rader2024optimistix}
Jason Rader, Terry Lyons, and Patrick Kidger.
\newblock Optimistix: modular optimisation in {JAX} and {Equinox}, 2024.
\newblock {arXiv}:2402.09983,
  \url{https://github.com/patrick-kidger/optimistix}.

\bibitem[Raue et~al.(2009)Raue, Kreutz, Maiwald, Bachmann, Schilling,
  Klingm{\"u}ller, and Timmer]{raue2009identifiability}
A.~Raue, C.~Kreutz, T.~Maiwald, J.~Bachmann, M.~Schilling, U.~Klingm{\"u}ller,
  and J.~Timmer.
\newblock Structural and practical identifiability analysis of partially
  observed dynamical models by exploiting the profile likelihood.
\newblock \emph{Bioinformatics}, 25\penalty0 (15):\penalty0 1923--1929, 2009.
\newblock \doi{10.1093/bioinformatics/btp358}.

\bibitem[Robertson(1966)]{robertson1966solution}
H.~H. Robertson.
\newblock The solution of a set of reaction rate equations.
\newblock In J.~Walsh, editor, \emph{Numerical Analysis: An Introduction},
  pages 178--182. Academic Press, London, 1966.

\bibitem[Rosenbrock(1963)]{rosenbrock1963general}
H.~H. Rosenbrock.
\newblock Some general implicit processes for the numerical solution of
  differential equations.
\newblock \emph{The Computer Journal}, 5\penalty0 (4):\penalty0 329--330, 1963.
\newblock \doi{10.1093/comjnl/5.4.329}.

\bibitem[Sandu(2006)]{sandu2006properties}
Adrian Sandu.
\newblock On the properties of {Runge--Kutta} discrete adjoints.
\newblock In \emph{Computational Science -- ICCS 2006}, volume 3994 of
  \emph{Lecture Notes in Computer Science}, pages 550--557. Springer, 2006.
\newblock \doi{10.1007/11758549_76}.

\bibitem[Sch{\"a}fer(1975)]{schafer1975hires}
Eberhard Sch{\"a}fer.
\newblock A new approach to explain the ``high irradiance responses'' of
  photomorphogenesis on the basis of phytochrome.
\newblock \emph{Journal of Mathematical Biology}, 2\penalty0 (1):\penalty0
  41--56, 1975.
\newblock \doi{10.1007/BF00276015}.

\bibitem[Serban and Hindmarsh(2005)]{serban2005cvodes}
Radu Serban and Alan~C. Hindmarsh.
\newblock {CVODES}: The sensitivity-enabled {ODE} solver in {SUNDIALS}.
\newblock In \emph{ASME International Design Engineering Technical Conferences
  and Computers and Information in Engineering Conference (IDETC/CIE)}, pages
  257--269, 2005.
\newblock \doi{10.1115/DETC2005-85597}.

\bibitem[Shampine and Reichelt(1997)]{shampine1997matlab}
Lawrence~F. Shampine and Mark~W. Reichelt.
\newblock The {MATLAB} {ODE} suite.
\newblock \emph{SIAM Journal on Scientific Computing}, 18\penalty0
  (1):\penalty0 1--22, 1997.
\newblock \doi{10.1137/S1064827594276424}.

\bibitem[Smith(2013)]{smith2013uncertainty}
Ralph~C. Smith.
\newblock \emph{Uncertainty Quantification: Theory, Implementation, and
  Applications}.
\newblock Society for Industrial and Applied Mathematics, Philadelphia, 2013.
\newblock \doi{10.1137/1.9781611973228}.

\bibitem[S{\"o}derlind(2002)]{soderlind2002automatic}
Gustaf S{\"o}derlind.
\newblock Automatic control and adaptive time-stepping.
\newblock \emph{Numerical Algorithms}, 31\penalty0 (1-4):\penalty0 281--310,
  2002.
\newblock \doi{10.1023/A:1021160023092}.

\bibitem[Steinebach(2023)]{steinebach2023rodas5p}
Gerd Steinebach.
\newblock Construction of {Rosenbrock--Wanner} method {Rodas5P} and numerical
  benchmarks within the {Julia} differential equations package.
\newblock \emph{BIT Numerical Mathematics}, 63\penalty0 (2):\penalty0 27, 2023.
\newblock \doi{10.1007/s10543-023-00967-x}.

\bibitem[Stumm and Walther(2010)]{stumm2010new}
Philipp Stumm and Andrea Walther.
\newblock New algorithms for optimal online checkpointing.
\newblock \emph{SIAM Journal on Scientific Computing}, 32\penalty0
  (2):\penalty0 836--854, 2010.
\newblock \doi{10.1137/080742439}.

\bibitem[Tarantola(2005)]{tarantola2005inverse}
Albert Tarantola.
\newblock \emph{Inverse Problem Theory and Methods for Model Parameter
  Estimation}.
\newblock Society for Industrial and Applied Mathematics, Philadelphia, 2005.
\newblock \doi{10.1137/1.9780898717921}.

\bibitem[Tsitouras(2011)]{tsitouras2011runge}
Ch. Tsitouras.
\newblock {Runge--Kutta} pairs of order 5(4) satisfying only the first column
  simplifying assumptions.
\newblock \emph{Computers \& Mathematics with Applications}, 62\penalty0
  (2):\penalty0 770--775, 2011.
\newblock \doi{10.1016/j.camwa.2011.06.002}.

\bibitem[Utkarsh et~al.(2024)Utkarsh, Churavy, Ma, Besard, Srisuma, Gymnich,
  Gerlach, Edelman, Barbastathis, Braatz, and Rackauckas]{utkarsh2024diffeqgpu}
Utkarsh Utkarsh, Valentin Churavy, Yingbo Ma, Tim Besard, Prakitr Srisuma, Tim
  Gymnich, Adam~R. Gerlach, Alan Edelman, George Barbastathis, Richard~D.
  Braatz, and Christopher Rackauckas.
\newblock Automated translation and accelerated solving of differential
  equations on multiple {GPU} platforms.
\newblock \emph{Computer Methods in Applied Mechanics and Engineering},
  419:\penalty0 116591, 2024.
\newblock \doi{10.1016/j.cma.2023.116591}.
\newblock {DiffEqGPU.jl}, \url{https://github.com/SciML/DiffEqGPU.jl}.

\bibitem[van~der Pol(1926)]{vanderpol1926relaxation}
Balthasar van~der Pol.
\newblock On ``relaxation-oscillations''.
\newblock \emph{The London, Edinburgh, and Dublin Philosophical Magazine and
  Journal of Science}, 2\penalty0 (11):\penalty0 978--992, 1926.
\newblock \doi{10.1080/14786442608564127}.

\bibitem[Verner(2010)]{verner2010numerically}
J.~H. Verner.
\newblock Numerically optimal {Runge--Kutta} pairs with interpolants.
\newblock \emph{Numerical Algorithms}, 53\penalty0 (2-3):\penalty0 383--396,
  2010.
\newblock \doi{10.1007/s11075-009-9290-3}.

\bibitem[Virtanen et~al.(2020)Virtanen, Gommers, Oliphant, Haberland, Reddy,
  Cournapeau, Burovski, Peterson, Weckesser, Bright, van~der Walt, Brett,
  Wilson, Millman, Mayorov, Nelson, Jones, Kern, Larson, Carey, Polat, Feng,
  Moore, VanderPlas, Laxalde, Perktold, Cimrman, Henriksen, Quintero, Harris,
  Archibald, Ribeiro, Pedregosa, van Mulbregt, and {SciPy 1.0
  Contributors}]{virtanen2020scipy}
Pauli Virtanen, Ralf Gommers, Travis~E. Oliphant, Matt Haberland, Tyler Reddy,
  David Cournapeau, Evgeni Burovski, Pearu Peterson, Warren Weckesser, Jonathan
  Bright, St{\'e}fan~J. van~der Walt, Matthew Brett, Joshua Wilson, K.~Jarrod
  Millman, Nikolay Mayorov, Andrew R.~J. Nelson, Eric Jones, Robert Kern, Eric
  Larson, C~J Carey, {\.I}lhan Polat, Yu~Feng, Eric~W. Moore, Jake VanderPlas,
  Denis Laxalde, Josef Perktold, Robert Cimrman, Ian Henriksen, E.~A. Quintero,
  Charles~R. Harris, Anne~M. Archibald, Ant{\^o}nio~H. Ribeiro, Fabian
  Pedregosa, Paul van Mulbregt, and {SciPy 1.0 Contributors}.
\newblock {SciPy} 1.0: fundamental algorithms for scientific computing in
  {Python}.
\newblock \emph{Nature Methods}, 17\penalty0 (3):\penalty0 261--272, 2020.
\newblock \doi{10.1038/s41592-019-0686-2}.
\newblock \url{https://github.com/scipy/scipy}.

\bibitem[Wang et~al.(2009)Wang, Moin, and Iaccarino]{wang2009minimal}
Qiqi Wang, Parviz Moin, and Gianluca Iaccarino.
\newblock Minimal repetition dynamic checkpointing algorithm for unsteady
  adjoint calculation.
\newblock \emph{SIAM Journal on Scientific Computing}, 31\penalty0
  (4):\penalty0 2549--2567, 2009.
\newblock \doi{10.1137/080727890}.

\bibitem[Zhang and Sandu(2014)]{zhang2014fatode}
Hong Zhang and Adrian Sandu.
\newblock {FATODE}: A library for forward, adjoint, and tangent linear
  integration of {ODEs}.
\newblock \emph{SIAM Journal on Scientific Computing}, 36\penalty0
  (5):\penalty0 C504--C523, 2014.
\newblock \doi{10.1137/130912335}.

\bibitem[Zhang et~al.(2022)Zhang, Constantinescu, and
  Smith]{zhang2022tsadjoint}
Hong Zhang, Emil~M. Constantinescu, and Barry~F. Smith.
\newblock {PETSc TSAdjoint}: A discrete adjoint {ODE} solver for first-order
  and second-order sensitivity analysis.
\newblock \emph{SIAM Journal on Scientific Computing}, 44\penalty0
  (1):\penalty0 C1--C24, 2022.
\newblock \doi{10.1137/21M140078X}.

\bibitem[Zhuang et~al.(2020)Zhuang, Dvornek, Li, Tatikonda, Papademetris, and
  Duncan]{zhuang2020aca}
Juntang Zhuang, Nicha Dvornek, Xiaoxiao Li, Sekhar Tatikonda, Xenophon
  Papademetris, and James Duncan.
\newblock Adaptive checkpoint adjoint method for gradient estimation in neural
  {ODE}.
\newblock In \emph{Proceedings of the 37th International Conference on Machine
  Learning (ICML)}, volume 119 of \emph{Proceedings of Machine Learning
  Research}, pages 11639--11649, 2020.
\newblock \doi{10.48550/arXiv.2006.02493}.
\newblock {arXiv}:2006.02493.

\bibitem[Zhuang et~al.(2021)Zhuang, Dvornek, Tatikonda, and
  Duncan]{zhuang2021mali}
Juntang Zhuang, Nicha~C. Dvornek, Sekhar Tatikonda, and James~S. Duncan.
\newblock {MALI}: A memory efficient and reverse accurate integrator for neural
  {ODEs}.
\newblock In \emph{International Conference on Learning Representations
  (ICLR)}, 2021.
\newblock \doi{10.48550/arXiv.2102.04668}.
\newblock {arXiv}:2102.04668.

\end{thebibliography}

\end{document}